\documentclass{natureprintstyle}
\usepackage{amssymb}
\usepackage{mathptmx}

\usepackage{astjnlabbrev-nature}

\usepackage{epsfig}
\usepackage{color}
\usepackage{graphicx}
\usepackage{longtable}

\newcommand{\spitzer}{{\it Spitzer}}
\newcommand{\chandra}{{\it Chandra}}
\newcommand{\hst}{{\it HST}}
\newcommand{\wise}{{\it WISE}}
\newcommand{\galex}{{\it GALEX}}

\newcommand{\herschel}{{\it Herschel}}
\newcommand{\xmm}{{\it XMM-Newton}}
\newcommand{\galfit}{{\sc galfit}}
\newcommand{\magphys}{{\sc magphys}}
\newcommand{\um}{$\mu$m}
\newcommand{\uJy}{$\mu$Jy}
\newcommand{\ergs}{{erg\,s$^{-1}$}}
\newcommand{\ergscm}{{erg\,s$^{-1}$\,cm$^{-2}$}}
\newcommand{\kms}{{km\,s$^{-1}$}}
\newcommand{\msun}{${\rm M}_{\odot}$}

\newcommand{\lsun}{${\rm L}_{\odot}$}
\newcommand{\msunyr}{${\rm M}_{\odot}~{\rm yr}^{-1}$}
\newcommand{\lir}{$L_{\rm IR}$}

\newcommand{\ms}{$M_{\rm star}$}
\newcommand{\mg}{$M_{\rm gas}$}
\newcommand{\mh}{$M_{\rm halo}$}

\newcommand{\mui}{$\mu^{-1}$}
\newcommand{\cohh}{CO$\to$H$_2$}
\newcommand{\arcsec}{$''$}
\newcommand{\arcmin}{$'$}
\newcommand{\nodata}{...}                                                                                                                 
\newcommand{\nd}{\nodata}
\newcommand{\obj}{HXMM01}
\newcommand{\objN}{X01N}
\newcommand{\objS}{X01S}

\newcommand{\sersic}{S\'{e}rsic}
\newcommand{\J}{$J$}
\newcommand{\K}{$K_{\rm S}$}
\newcommand{\co}{CO~$J$~=~1$\to$0}
\newcommand{\cothree}{CO~$J$~=~3$\to$2} 
\newcommand{\cofour}{CO~$J$~=~4$\to$3} 
\newcommand{\citep}{\cite}
\newcommand{\citet}{\cite}

\usepackage{longtable}
\usepackage[labelsep=endash]{caption}

\title{The rapid assembly of an elliptical galaxy of 400 billion solar masses at a redshift of 2.3}

\author{Hai~Fu$^{1}$, Asantha~Cooray$^{1}$, C.~Feruglio$^{2}$, R.J.~Ivison$^{3}$, D.A.~Riechers$^{4}$, M.~Gurwell$^{5}$, R.S.~Bussmann$^{5}$, A.I.~Harris$^{6}$, B.~Altieri$^{7}$, H.~Aussel$^{8}$, A.J.~Baker$^{9}$, J.~Bock$^{10,11}$, M.~Boylan-Kolchin$^{1}$, C.~Bridge$^{11}$, J.A.~Calanog$^{1}$, C.M.~Casey$^{12}$, A.~Cava$^{13}$, S.C.~Chapman$^{14}$, D.L.~Clements$^{15}$, A.~Conley$^{16}$, P.~Cox$^{2}$, D.~Farrah$^{17}$, D.~Frayer$^{18}$, R.~Hopwood$^{15}$, J.~Jia$^{1}$, G.~Magdis$^{19}$, G.~Marsden$^{20}$, P.~Mart{\'i}nez-Navajas$^{21,22}$, M.~Negrello$^{23}$, R.~Neri$^{2}$, S.J.~Oliver$^{24}$, A.~Omont$^{25}$, M.J.~Page$^{26}$, I.~P{\'e}rez-Fournon$^{21,22}$, B.~Schulz$^{27}$, D.~Scott$^{20}$, A.~Smith$^{24}$, M.~Vaccari$^{28}$, I.~Valtchanov$^{7}$, J.D.~Vieira$^{11}$, M.~Viero$^{11}$, L.~Wang$^{24}$, J.L.~Wardlow$^{1}$, M.~Zemcov$^{10}$}
\begin{document}
\maketitle

\begin{abstract}

Stellar archeology\cite{McCarthy04} shows that massive elliptical galaxies today formed rapidly 
about ten billion years ago with star formation rates above several 
hundreds solar masses per year (M$_\mathbf{\odot}$~yr$^\mathbf{-1}$). 
Their progenitors are likely the sub-millimeter-bright galaxies (SMGs)\cite{Barger98} 
at redshifts ($\mathbf{z}$) greater than 2. 
While SMGs' mean molecular gas mass\cite{Bothwell12} of $\mathbf{5\times10^{10}}$ M$_\mathbf{\odot}$ 
can explain the formation of typical elliptical galaxies, it is inadequate to form ellipticals\cite{Damjanov09} that 
already have stellar masses above $\mathbf{2\times10^{11}}$ M$_\mathbf{\odot}$ at $\mathbf{z \approx 2}$. 
Here we report multi-wavelength high-resolution observations of a rare merger 
of two massive SMGs at $\mathbf{z = 2.3}$. The system 
is currently forming stars at a tremendous rate of 2,000 M$_\mathbf{\odot}$~yr$^\mathbf{-1}$. 
With a star formation efficiency an order-of-magnitude greater than 
that of normal galaxies, it will quench the star formation by exhausting the 
gas reservoir in only $\mathbf{\sim200}$ million years. 
At a projected separation of 19~kiloparsecs, the two
massive starbursts are about to merge and form a passive elliptical galaxy 
with a stellar mass of $\mathbf{\sim4\times10^{11}}$ M$_\mathbf{\odot}$. 
Our observations show that gas-rich major galaxy mergers,
concurrent with intense star formation, can form the most
massive elliptical galaxies by $\mathbf{z \approx 1.5}$. 

\end{abstract}

\let\thefootnote\relax\footnote{
\begin{affiliations}
\item Department of Physics \& Astronomy, University of California, Irvine, CA 92697, USA
\item Institut de RadioAstronomie Millim\'etrique, 300 Rue de la Piscine, Domaine Universitaire, 38406 Saint Martin d'H\`eres, France
\item UK Astronomy Technology Centre, Royal Observatory, Edinburgh, EH9 3HJ, UK
\item Department of Astronomy, Cornell University, 610 Space Science Building, Ithaca, NY 14853, USA
\item Harvard-Smithsonian Center for Astrophysics, 60 Garden Street, Cambridge, MA 02138, USA
\item Department of Astronomy, University of Maryland, College Park, MD 20742-2421, USA
\item Herschel Science Centre, European Space Astronomy Centre, Villanueva de la Ca\~nada, 28691 Madrid, Spain
\item Laboratoire AIM-Paris-Saclay, CEA/DSM-CNRS-Universit\'e Paris Diderot, Irfu/SAp, CEA-Saclay, 91191 Gif-sur-Yvette Cedex, France
\item Department of Physics and Astronomy, Rutgers, The State University of New Jersey, 136 Frelinghuysen Rd, Piscataway, NJ 08854, USA
\item Jet Propulsion Laboratory, 4800 Oak Grove Drive, Pasadena, CA 91109, USA
\item California Institute of Technology, 1200 E. California Blvd., Pasadena, CA 91125, USA
\item Institute for Astronomy, University of Hawaii, 2680 Woodlawn Dr., Honolulu, HI 96822, USA
\item Departamento de Astrof\'isica, Facultad de CC. F\'isicas, Universidad Complutense de Madrid, E-28040 Madrid, Spain
\item Department of Physics \& Atmospheric Science, Dalhousie University, 6310 Coburg Road, Halifax B3H 4R2, Canada
\item Astrophysics Group, Imperial College London, Blackett Laboratory, Prince Consort Road, London SW7 2AZ, UK
\item Center for Astrophysics and Space Astronomy 389-UCB, University of Colorado, Boulder, CO 80309, USA
\item Department of Physics, Virginia Tech, Blacksburg, VA 24061, USA
\item NRAO, PO Box 2, Green Bank, WV 24944, USA
\item Department of Physics, University of Oxford, Keble Road, Oxford OX1 3RH, UK
\item Department of Physics \& Astronomy, University of British Columbia, 6224 Agricultural Road, Vancouver, BC V6T~1Z1, Canada
\item Instituto de Astrof{\'\i}sica de Canarias (IAC), E-38200 La Laguna, Tenerife, Spain
\item Departamento de Astrof{\'\i}sica, Universidad de La Laguna (ULL), E-38205 La Laguna, Tenerife, Spain
\item INAF - Osservatorio Astronomico di Padova, Vicolo dell'Osservatorio 5, I-35122 Padova, Italy
\item Astronomy Centre, Dept. of Physics \& Astronomy, University of Sussex, Brighton BN1 9QH, UK
\item Institut d'Astrophysique de Paris, UMR 7095, CNRS, UPMC Univ. Paris 06, 98bis boulevard Arago, F-75014 Paris, France
\item Mullard Space Science Laboratory, University College London, Holmbury St. Mary, Dorking, Surrey RH5 6NT, UK
\item Infrared Processing and Analysis Center, MS 100-22, California Institute of Technology, JPL, Pasadena, CA 91125, USA
\item Astrophysics Group, Physics Department, University of the Western Cape, Private Bag X17, 7535, Bellville, Cape Town, South Africa
\end{affiliations}
}

%%
%% Figures
%%
\begin{figure*}
\centerline{
\includegraphics[width=0.85\textwidth]{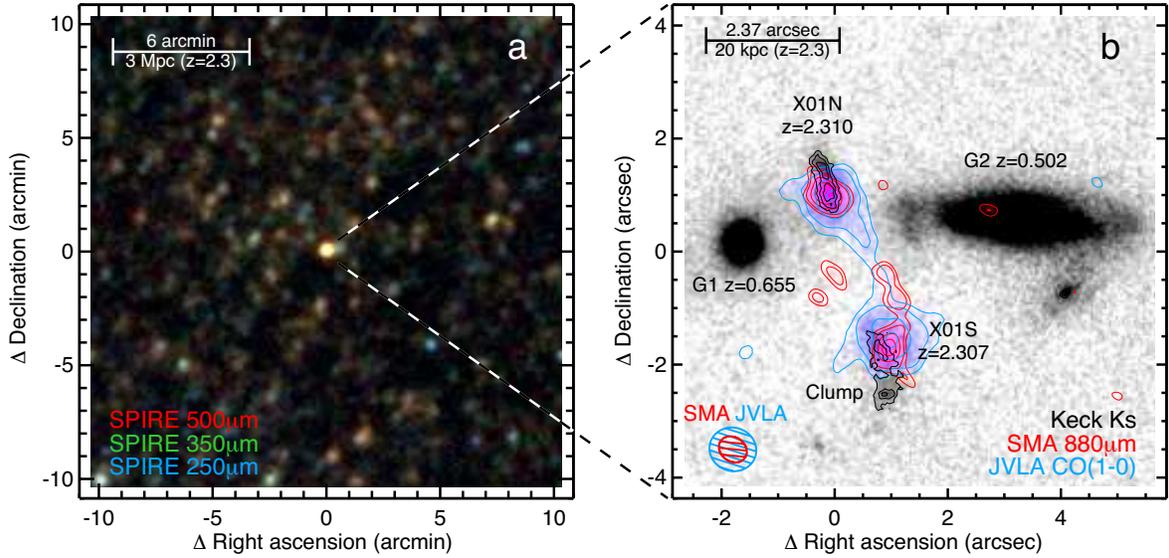}
}
\caption{\textbf{Multi-wavelength view of \obj}. Panel (a) shows the
\herschel\ Space Observatory's false three-color image combining 250
(blue), 350 (green) and 500 (red) $\mu$m images with resolutions of 18,
25 and 36\arcsec, respectively. \obj\ is the brightest source in the
image.  Panel (b) shows the highest resolution images of \obj. The
background and black contours show the Keck-NIRC2 \K-band adaptive
optics image. Overlain are the dust continuum emission at 880\,\um\ from
the Submillimeter Array (SMA; red contours) and the molecular \co\ emission from the Jansky Very Large Array 
(JVLA; blue contours).  The SMA and JVLA contours are drawn at +3, +4, +6,
and +8$\times$ the rms noise ($\sigma$ = 0.67\,mJy\,beam$^{-1}$ for SMA,
and $\sigma$ = 19\,\uJy\,beam$^{-1}$ for JVLA), and the Keck contours
are at +5, +8, +11$\sigma$, where $\sigma$ is again the rms noise
($3.7\times10^{-3}$~\uJy~pixel$^{-1}$). The two major components of
\obj\ (\objN\ and \objS) and the foreground galaxies are labelled along
with their redshifts. Note that there is also a bridge of material
detected at $\gtrsim$5$\sigma$ level between \objN\ and \objS\ in the
SMA and JVLA images. We also label the southern part of \objS\ as a
``clump'' because of its distinct optical spectral energy distribution 
compared to the rest of \obj. This clump could be a contaminating 
source, so we have excluded it
in our stellar mass estimate of \objS. The ellipses at the lower left show the
beam full width half maximum (0.54\arcsec$\times$0.44\arcsec\ for the
SMA and 0.83\arcsec$\times$0.77\arcsec\ for the JVLA).  \label{fig:obs}}
\end{figure*}

\begin{figure*}
\centerline{
\includegraphics[width=1.0\textwidth]{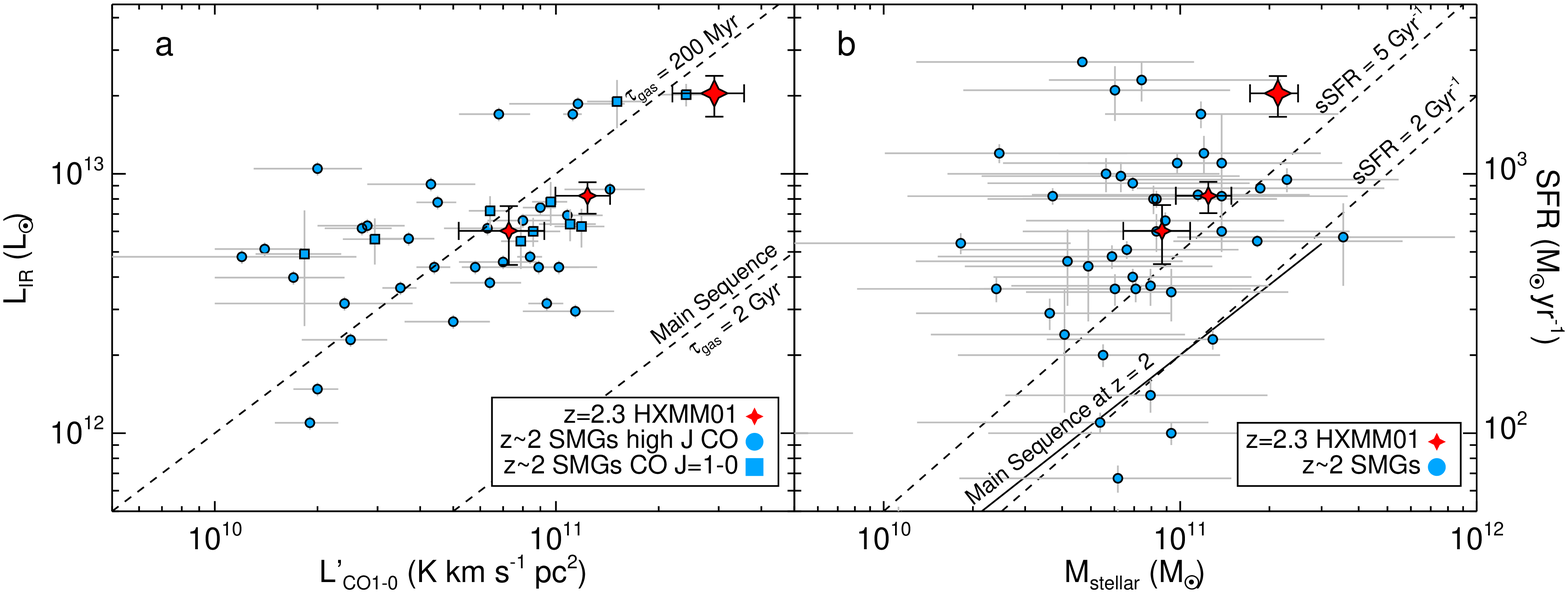}
}
\caption{\textbf{The infrared luminosity and stellar masses of sub-millimeter galaxies.}
In both panels \obj\ is plotted as the
large filled red star. Its two main
components, \objN\ and \objS, are plotted separately as the small
filled red stars.  The panel (a) plots IR luminosity vs. \co\ luminosity with
blue squares showing SMGs with direct \co\ measurements\cite{Riechers11b,Ivison11}.
The blue circles are SMGs with higher $J$ CO line luminosities
that have been converted to \co\ luminosities using the mean observed ratios\cite{Bothwell12}.
Assuming $\alpha_{\rm CO} = 1$, the dashed lines indicate constant consumption timescales of the entire gas reservoir 
($\tau_{\rm gas} \equiv$ 2 \mg/SFR) of 200~Myr and 2~Gyr, which are the mean values for
SMGs\cite{Bothwell12} and massive (\ms~$> 10^{11}$~\msun) normal
star-forming galaxies on the main sequence\cite{Genzel12} at $z \sim
2$, respectively.  The panel (b) plots SFR vs. stellar mass with
blue data points showing SMGs at $z \sim 2$ using data
from the literature\cite{Bothwell12,Riechers11b,Ivison11,Hainline08}. 
The star-forming main sequence\cite{Daddi07} at $z = 2$ is indicated by a solid line. Dashed
lines show constant specific SFRs (sSFR~$\equiv$~SFR/\ms). We find a
sSFR for \obj\ of $9.6\pm2.3$~Gyr$^{-1}$. At $z \sim 2$ this places
\obj\ five times higher than the main sequence of normal
star-forming galaxies\cite{Elbaz11}.  All error bars are formal $\pm$1$\sigma$
standard deviations, including those associated with the lensing magnification. 
The two panels demonstrate that \obj\ has one of the largest gas and stellar content when compared to the SMG population at
$z \sim 2$, and it is clearly in a starburst phase.  
\label{fig:lirlco}}
\end{figure*}

\begin{figure}
\centerline{
\includegraphics[width=0.5\textwidth]{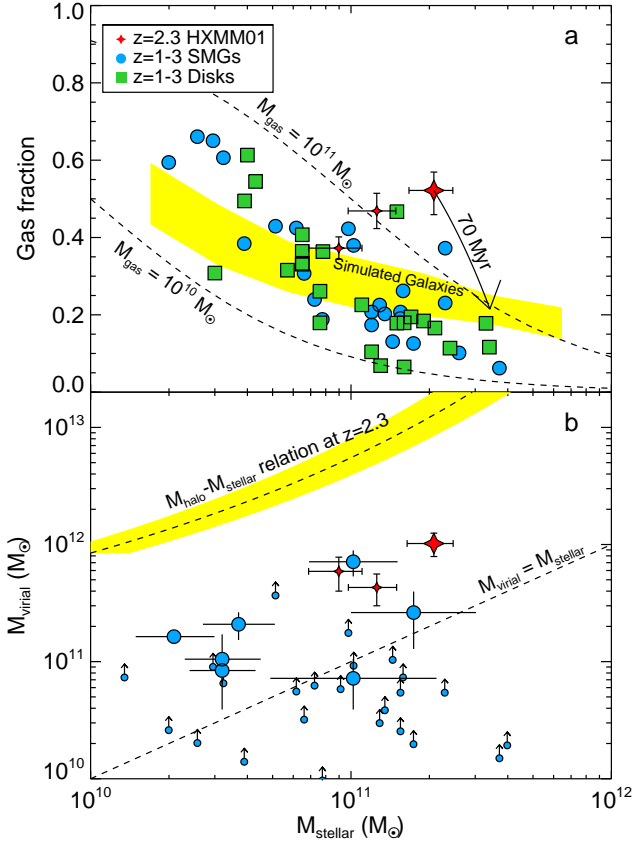}
}
\caption{\textbf{The gas mass and dynamical state of sub-millimeter galaxies.}
The panel (a) shows the gas-to-baryon fraction vs. stellar mass for
star-forming galaxies. \obj\ is plotted as the large filled red star, 
while its two main components, \objN\ and \objS, are the two small filled red
stars. Because a significant fraction of \co\ emission is detected
outside of the two main components, the total gas fraction is higher
than those of \objN\ and \objS. In panel (a), the other data points
are for SMGs (blue circles) and main-sequence star-forming galaxies (green squares). Both samples have used gas masses estimated from a model-calibrated relation
between $\alpha_{\rm CO}$ and CO brightness temperature\cite{Narayanan12b}.  The same relation is used to estimate
$\alpha_{\rm CO}$ for \obj. The yellow stripe shows $z \sim 2$
star-forming galaxies from cosmological hydrodynamic
simulations\cite{Dave10}. The solid curve with an arrow shows the
expected evolution of \obj\ in the next 70~Myr assuming the observed
rate of star formation and the conservation of mass.
The dashed curves indicate constant gas masses of $10^{10}$ and
$10^{11}$~\msun. The  panel (b) shows the virial dynamical mass vs. stellar mass
for SMGs at $\mathbf{z \sim 2}$ with bigger and smaller blue circles showing
SMGs from two other studies\cite{Engel10,Bothwell12}.  The yellow
stripe shows the stellar-mass$-$halo-mass relation at $z = 2.3$ from
abundance matching\cite{Behroozi10}. All error bars are formal
$\pm$1$\sigma$ standard deviation, including those associated with the
lensing magnifications. \obj\ is extremely gas-rich compared to other $z
\sim 2$ star-forming galaxies with similar stellar masses. It
represents a galaxy evolution stage that is difficult to reproduce in
simulations. The stellar mass of \obj\ also implies a halo mass of
$\sim5\times10^{12}$~\msun\ for each of the merging galaxies; i.e., a
merged halo mass of $10^{13}$~\msun.  \label{fig:fgas}} 
\end{figure}
%%%%%%%%%%%%%%%%%%%%%%%%%%%%%%%%%

%%
%% Tables
%%

\begin{table*}
\begin{center}
\begin{tabular}{lccccc}
\hline
\hline
Object & z & $S_{850}$ & Sep & $M_{\rm H2}$ & SFR \\
       &     & (mJy) & (kpc) & ($10^{10}$\,$M_{\odot}$) & ($M_{\odot}~yr^{-1}$) \\
\hline
SMM~J02399$-$0136   & 2.8 & 9    & 8  & 8.0  & 990 \\
SMM~J09431$+$4700   & 3.4 & 10.5 & 30 & 6.6 & 1160 \\
SMM~J105141         & 1.2 & 8.7  & 5  & 6.6 & 960 \\
SMM~J123707/HDF~242 & 2.5 & 4.7  & 22 & 7.4 & 520 \\
SMM~J123711/HDF~254 & 2.0 & 4.2  & 8  & 0.8 & 650 \\
SMM~163650/N2~850.4 & 2.4 & 8    &  4 & 9.6  & 880 \\
4C~60.07            & 3.8 & 11   & 25 & 6.4 & 1200 \\
\hline
\obj                & 2.3 & 20$\pm$4 & 19$\pm$4 & 23$\pm$6 & 2000$\pm$400 \\
\hline
\end{tabular}
\end{center}
\caption{{\bf Resolved sub-millimeter galaxy mergers with projected separations less than 30~kpc}.
The molecular gas masses ($M_{\rm H2}$) are estimated with the conversion factor from \co\
luminosity to molecular gas mass, $\alpha_{\rm CO}$.  This factor is
estimated to be about four for normal star-forming galaxies like
the Milky Way\cite{Strong96} and about one for local starbursts\cite{Downes98}. 
Two independent methods outlined in the Supplementary Information, using either the gas-to-dust mass ratio or
the velocity-integrated \co\ brightness temperature\cite{Narayanan12a}, lead to
$\alpha_{\rm CO} \sim 0.8$~\msun/(K~km~s$^{-1}$~pc$^{-2}$) for \obj,
consistent with the $\alpha_{\rm CO}$ measurements for several other
SMGs\cite{Magdis11,Magnelli12b}. Using this conversion factor,
we find that \obj\ contains $(2.3\pm0.6)\times10^{11}$~\msun\ of molecular gas.
We use the same $\alpha_{\rm CO} = 0.8$ for other SMGs and have used \co\ measurements wherever available. 
SFRs are estimated using the 850~\um\ flux density (SFR/\msunyr\ = 110 $S_{850}$/mJy) and are uncertain by $\lesssim$50\%. For \obj, this relation gives a SFR consistent with 
that from  the total IR luminosity. Similar to \obj\ values for SMM~J02399$-$0136 have been corrected for lensing magnification. 
}
\label{tab:smgmerger}
\end{table*}
%%%%%%%%%%%%%%%%%%%%%%%%%%%%

\obj\ (1HERMES\,S250\,J022016.5$-$060143) was identified as an unusually bright
SMG\cite{Wardlow13} in the \herschel\ Multi-tiered Extragalactic Survey (HerMES\cite{Oliver12}). 
We observed it with a variety of ground-based
telescopes from optical to radio to obtain higher resolution
images and to better sample the spectral energy distribution (SED). 
The source is resolved into two similarly bright components at $z \sim 2.308$ separated by
3\arcsec\ (Figure~\ref{fig:obs}; hereafter
\objN\ and \objS\ for the northern and southern component, respectively), 
which are connected by a bridge of material reminiscent of tidal tails
frequently seen in galaxy mergers. The Carbon Monoxide (CO) $J$~=~1$\to$0, $J$~=~4$\to$3, and
H$\alpha$ spectra all show slightly different redshifts for these two
components. The velocity separation is $260\pm70$~\kms\ and $330\pm220$~\kms\ from
the \co\ and H$\alpha$ spectra, respectively. In addition, the \co\ line
widths for the two components are significantly different
($970\pm150$~\kms\ vs. $660\pm100$~\kms), establishing clearly that they
are two distinct galaxies undergoing a merger. \obj\ is close to two low-redshift galaxies 
and is weakly gravitationally magnified.
From our lens model (see Supplementary Information), we determined that the magnification factors are 
$1.8\pm0.5$, $1.4\pm0.2$, and $1.6\pm0.3$ for \objN, \objS, and the
entire system, respectively. Here we quote lensing-corrected values, and 
the uncertainties in magnification have been propagated into their errors.

With a flux density of $20\pm4$~mJy at 850~\um, \obj\ is among the brightest 
SMGs known, which typically 
have a flux density of 5~mJy at 850~\um\ of typical SMGs. The SED between
160~\um\ and 2.1~mm is fully consistent with a modified blackbody with a
characteristic dust temperature of $55\pm3$~K. The dust temperature is
much hotter than those of normal star-forming galaxies, and lies at the 
higher end of starbursts\cite{Magnelli12b}. We determine a total infrared
(IR; 8$-$1000~\um) luminosity of \lir~$= (2.0\pm0.4)\times10^{13}$~\lsun\ 
for all components combined, where \lsun\ is the solar luminosity. This luminosity 
implies an instantaneous star formation
rate\cite{Kennicutt98b} (SFR) of $2000 \pm 400$~\msunyr\ for a Chabrier
initial mass function. 

We estimate the total molecular gas mass from the \co\ line luminosity. 
The area-velocity-integrated \co\ brightness temperature, which is also 
a luminosity measure, is
$(2.9\pm0.7)\times10^{11}$~K~km~s$^{-1}$~pc$^{2}$, implying a mass of 
$(2.3\pm0.6)\times10^{11}$~\msun\ of molecular gas using our derived 
conversion factor (see Supplementary Information). 
This is the highest intrinsic \co\ line luminosities among SMGs in the 
literature (Figure~\ref{fig:lirlco}).
Table~\ref{tab:smgmerger} lists the known SMG mergers that are well separated in to two galaxies. 
It shows that \obj\ is the brightest, most luminous, and most gas-rich SMG merger that is
known.

We estimate stellar masses of $(9 \pm 2)\times10^{10}$\,\msun\ and $(1.3
\pm 0.3)\times10^{11}$\,\msun\ for \objN\ and \objS\ respectively, by
modeling their UV-to-millimeter SEDs using the public code
MAGPHYS\cite{Cunha08}. \obj\ is hence a merger of two massive galaxies
in terms of both stellar and gas content. The combined gas-to-baryon
fraction of \obj\ is $(52 \pm 5)$\%, far above the mean gas-to-baryon
fraction of star-forming galaxies with stellar masses
above $10^{11}$~\msun\ at $z \sim 2$ (Figure~\ref{fig:fgas}). Massive
galaxies with such a high gas fraction are not reproduced in cosmological
simulations\cite{Dave10}. \obj\ may represent a system that has accumulated 
a disproportionate amount of gas because of 
a higher accretion rate compared to the SFR in the past. Assuming a 
constant SFR with the observed value and no additional gas accretion, 
\obj\ will reach a gas fraction of $\sim$20\% in just $\sim$70~Myr,
such that its stellar mass and gas fraction will be 
consistent with the simulated galaxies.

Because the components of \obj\ are resolved in our SMA and JVLA
observations, we use them to directly measure the physical extent of the
dust and gas emission. We find that the intrinsic sizes of the dusty
star forming regions ($A_{880} = 5-7$\,kpc$^2$) are $3-7\times$ smaller
than the molecular gas reservoirs ($A_{\rm CO} = 15-50$~kpc$^2$). This
is in direct conflict with the commonly used assumption that the gas and
dust in SMGs have the same physical extent. This discrepancy has also
been observed in several other SMGs\cite{Ivison11}.  When the sizes are
used to measure to gas and star formation surface densities, we find
\obj\ lies along the sequence of local and high-redshift starbursts in
the Kennicutt-Schmidt relation\cite{Kennicutt98b} (see Supplementary
Information).  We also estimate the star formation efficiency,
$\epsilon_{\rm SF}$, the percentage of gas that is converted into stars
in a dynamical timescale.  We find that the two merging SMGs are
remarkably efficient in forming stars with values of $\epsilon_{\rm SF}
= (10 \pm 3)$\% and $(41 \pm 10)$\% for \objN\ and \objS, respectively.
Similar to other starbursts\cite{Genzel10}, these efficiencies are about
an order-of-magnitude higher than the typical value of normal
star-forming galaxies ($\epsilon_{\rm SF} \sim 2$\%). These efficiencies reach
the theoretical maximum of 30\% that has been estimated for
molecular cloud-dominated starbursts\cite{Murray10}.

\obj\ illustrates the rapid formation of an extremely massive elliptical galaxy 
in an equal mass gas-rich merger. The high star formation efficiency coupled 
with the short dynamical timescale would quickly turn this pair of starburst 
galaxies into a passive elliptical galaxy, as its SFR will decline with an 
$e$-folding time of $230\pm40$~Myr ($\tau_{\rm gas} \equiv 2 M_{\rm gas}/$SFR) 
even without any feedback process\cite{Di-Matteo05} ejecting gas out of the galaxy. 
We have assumed no additional gas infall. The factor of two in our
gas exhausting timescale ($\tau_{\rm gas}$) accounts for the 50\% gas recycling 
in stellar evolution. 
The descendent of \obj\ is expected to 
have a stellar mass of $\sim4\times10^{11}$~\msun, comparable to the most massive 
elliptical galaxies\cite{Damjanov09} at
$z \sim 2$. Its stellar population will appear old and passively evolving by $z \sim 1.7$ 
(i.e., 1~Gyr afterwards) because the star formation ceased so rapidly\cite{Martin07}. 
While gas-poor (``dry'') merging\cite{Dokkum05}
between less massive ellipticals are generally
invoked to explain the subsequent stellar mass buildup after the SMG
phase, the formation of a very massive galaxy directly through a
gas-rich merger shows that dry mergers are not the only channel to form 
the most massive ellipticals. They can form by {\it in situ} starbursts 
involved in massive SMG mergers similar to \obj.

Starburst mergers similar to \obj\ are extremely rare because they are short-lived and unusually massive. 
In addition, only a small fraction of mergers are expected to be observed as two separate galaxies\cite{Engel10}. 
Therefore, although we estimate that only one SMG$-$SMG merger 
as bright and luminous as \obj\ to be present in every $\sim$100~deg$^2$, their importance to galaxy evolution 
must be evaluated after correcting for their limited visibility due to the wide separation and the short lifetime. 
We can estimate the space density of massive mergers like \obj\ based on its CO luminosity, assume all 
SMGs with such high CO luminosities are also mergers.
But because the space density of galaxies as a function of CO luminosities  
is largely unknown, we resort to the observed space density of SMGs as a function 
of the SFR\cite{Chapman05}, given that SFR is proportional to gas mass. 
The space density of SMGs with SFR greater than 2000~\msunyr\ is $(2.4\pm1.2)\times10^{-6}$~Mpc$^{-3}$ 
at $2 < z < 3$. Give the time span covered by the redshift range is 1.2~Gyr and 
the lifetime of \obj\ is $\sim200$~Myr, the space density of \obj-like galaxies is $(1.4\pm0.7)\times10^{-5}$~Mpc$^{-3}$ 
after correcting for the 17\% duty cycle. Despite the large uncertainty, 
this is comparable to the space density of passive 
elliptical galaxies\cite{Ilbert10} with stellar masses above $2\times10^{11}$~\msun\ at $z \sim 1.1$.
Therefore, although SMGs as luminous as \obj\ are rare, they could represent a short but critical transitional phase 
in the early formation of the most massive elliptical galaxies. This conclusion will soon be tested 
with wide-area sub-millimeter surveys that will uncover more SMG$-$SMG mergers similar to \obj.\\

\noindent
\textbf{\textsf{\footnotesize Received 15 February; accepted 12 April 2013.}}

\begin{addendum} 

\item[Acknowledgements] 
\herschel\ is an ESA space observatory with science instruments provided
by European-led Principal Investigator consortia and with important
participation from NASA. This research has made use of data from the
HerMES project (http://hermes.sussex.ac.uk/). HerMES is a \herschel\ Key
Programme utilizing Guaranteed Time from the SPIRE instrument team, ESAC
scientists and a mission scientist. SPIRE has been developed by a
consortium of institutes led by Cardiff University (UK) and including
Univ. Lethbridge (Canada); NAOC (China); CEA, LAM (France); IFSI, Univ.
Padua (Italy); IAC (Spain); Stockholm Observatory (Sweden); Imperial
College London, RAL, UCL-MSSL, UKATC, Univ. Sussex (UK); and
Caltech/JPL, IPAC, Univ.  Colorado (USA). This development has been
supported by national funding agencies: CSA (Canada); NAOC (China); CEA,
CNES, CNRS (France); ASI (Italy); MCINN (Spain);  SNSB (Sweden); STFC
(UK); and NASA (USA).  The data presented in this paper will be released
through the HeDaM Database in Marseille at http://hedam.oamp.fr/HerMES.

\item[Author Contributions] 
H.F. and A.C. wrote the manuscript and had the overall lead of the project.
C.F., R.J.I., D.A.R., M.G., R.S.B., and A.I.H. contributed significantly 
to the taking and analysis of the follow-up data with various instruments.
All other coauthors of this paper contributed extensively and equally by
their varied contributions to the HerMES project, planning of HerMES
observations, analysis of HerMES data, and by commenting on this
manuscript as part of an internal review process.

\item[Correspondence] Correspondence and requests for materials should
be addressed to H.F. (haif@uci.edu) and A.C. (acooray@uci.edu)

\item[Competing interests statement] The authors declare no competing
interests.
 
\item[Supplementary information] accompanies this paper. 

\end{addendum}

\clearpage

\setcounter{page}{1}
\setcounter{figure}{0}
\setcounter{table}{0}
\renewcommand{\thefigure}{S\arabic{figure}}
\renewcommand{\thetable}{S\arabic{table}}

\begin{center}
{\bf \Large \uppercase{Supplementary Information} }
\end{center}

Here we give details of the multi-wavelength observations (\S~\ref{sec:obs}), the photometry deblending procedures (\S~\ref{sec:deblend}), the weak lensing model (\S~\ref{sec:lensing}), and the estimation of the physical properties (\S~\ref{sec:properties}) and the space density of \obj\ (\S~\ref{sec:density}).

Throughout we adopt a Chabrier\cite{Chabrier03}
initial mass function (IMF), and the concordance $\Lambda$CDM cosmology\cite{Komatsu11}
with $\Omega_{\rm m}=0.27$, $\Omega_\Lambda=0.73$, and $H_0$ = 70~km~s$^{-1}$~Mpc$^{-1}$.

\section{Observations and Data Reduction} \label{sec:obs}

\obj\ was initially identified as a strong lensing
candidate\cite{Wardlow13} in SPIRE\cite{Griffin10} observations from the \herschel\footnote{\herschel\ is an ESA space observatory with science instruments
provided by European-led Principal Investigator consortia and with
important participation from NASA.} Space Observatory\cite{Pilbratt10}
as part of the HerMES\footnote{hermes.sussex.ac.uk} survey\cite{Oliver12}
of the XMM-LSS field. The source has an unusually high 500\,\um\ flux density
($S_{500} = 132$\,mJy), which suggests
gravitational lensing of a fainter background galaxy\cite{Negrello10}.

Along with other lensing candidates\cite{Wardlow13}, we observed \obj\ with
the Submillimeter Array\cite{Ho04} to refine its position and to
examine if its morphology is consistent with gravitational lensing. The
\herschel\ source was resolved into two similarly bright SMA sources
separated by only $\sim$3\arcsec\ and close to two low-redshift
galaxies, making it a high-priority target for further followup
observations.  

Guided by the far-IR photometric
redshift, the spectroscopic redshift of \obj\ was first determined from
a blind search for \cothree\ emission with the Combined Array for
Research in Millimeter-wave Astronomy (CARMA). Subsequently, the
redshift was confirmed at IRAM/Plateau de Bure Interferometer (PdBI) with a measurement of
\cofour\  emission. We also obtained single-dish and interferometric \co\
data with the National Radio Astronomy Observatory\footnote{NRAO is
operated by Associated Universities Inc., under a cooperative agreement
with the National Science Foundation.}'s Robert C.\ Byrd Green Bank
Telescope (GBT) and Karl G. Jansky Very Large Array (JVLA)\cite{Perley11}.
Finally, with Keck, we obtained a near-IR spectrum covering the
H$\alpha$ emission of \obj\ and optical spectra of the nearby foreground
galaxies to determine their redshifts. The Keck \K-band adaptive optics image and the \hst/WFC3
F110W image further revealed the clumpy rest-frame optical
morphology of both SMA sources. In the following, we describe the
observations and data reduction procedures in more details.

\subsection{Herschel Far-IR Imaging} \label{sec:herschel}

The original \herschel\ SPIRE imaging data for this source, as part of the HerMES XMM-LSS field observations\cite{Oliver12}, are now publicly available from the \herschel\ Science Archive. The observational identifications (Obs Ids) of those datasets are 1342189003 and 1342190312. As the parallel-mode Photodetector Array Camera and Spectrometer (PACS)\cite{Poglitsch10} data at 100 and 160~\um\ is shallow, we then obtained deeper scans of \obj\ with integration times of 840~s at these wavelengths using part of the remaining SPIRE team's guaranteed time (Obs Ids: 1342258806 and 1342258807). Finally, a 70~\um\ scan map of \obj\ with an integration time of 276~s was obtained as part of an OT2 open-time program (Obs Ids: 1342261958 and 1342261957). The SPIRE data are reduced and analyzed following the standard procedures\cite{Levenson10}. We used the high fidelity maps produced using an internal map-making package\cite{Viero12}. The PACS data are reduced with HIPE.

\subsection{Submillimeter Imaging} \label{sec:sma}
 
We obtained high-resolution interferometric imaging of \obj\ at 876.5$\,\mu$m
(342.02~GHz) with the SMA (PIs: Gurwell, Bussmann). \obj\ was observed in the subcompact array 
configuration on 2010 Aug 14 with an on-source integration time ($t_{\rm int}$) of
58~min, in the extended configuration on 2010 Sep 26 with $t_{\rm int} =
141$~min, and in the very extended configuration on 2011 Jan 4 with $t_{\rm int} = 54$~min. Atmospheric opacity
and phase noise for all observations was very low ($\tau_{\rm 225~GHz} = 0.07$,
$\phi \approx 20$~deg).  All observations used an intermediate frequency (IF)
coverage of 4Ð8 GHz to provide a total of 8~GHz bandwidth (considering both
sidebands). The quasars 0217+017 and 0238+116 were used for time-variable gain
(amplitude and phase) calibration. 3C~454.3 served as the primary bandpass
calibrator.  We used a combination of Uranus and Callisto as the absolute flux
calibrators.  

We used the MIR software package in Interactive Data Language (IDL) to calibrate the visibility data.  For
imaging, we used the Multichannel Image Reconstruction, Image Analysis, and
Display (MIRIAD) software package with natural weighting for maximum
sensitivity.  The final CLEANed image has a synthesized beam with a FWHM
resolution of 0.52\arcsec$\times$0.43\arcsec\ at a position angle (PA) of $55^{\circ}$ east of north
and an root-mean-square noise level of 0.67~mJy~beam$^{-1}$. The primary
beam of the SMA is 35\arcsec\ FWHM at 342~GHz.

The final SMA map resolved \obj\ into two major components separated by
2.95\arcsec\ (25~kpc) along with filamentary structures connecting the
two (Figure~\ref{fig:obs}). The northern component (\objN) has a flux density of $9.3\pm0.7$\,mJy, while the southern component (\objS) has a flux density of $8.3\pm0.7$\,mJy. The filaments have a total flux density of $9.4\pm1.0$\,mJy, accounting for $\sim$35\% of the entire system ($27.0\pm1.4$\,mJy).

\subsection{Near-IR Imaging} \label{sec:nir}

We obtained 40$\times$80\,s \K-band images on 2012 February 4 (UT) with
the Keck\,II laser guide-star adaptive-optics system\cite{Wizinowich06} (PI: Cooray).
We used the NIRC2 camera at 0.04\arcsec~pixel$^{-1}$ scale (40\arcsec\
field), and we dithered with 2$-$3\arcsec\ steps in a nine-point
dithering pattern. The atmospheric seeing at 0.5\,\um\ was
$\sim$0.8\arcsec. An $R = 17.6$ magnitude star 67\arcsec\ SW of \obj\
served as the tip-tilt reference star. The estimated Strehl ratio at
the source position is $\sim$14\% under normal conditions. Because of
the large separation from the tip-tilt star, we had to rotate the camera
to PA = 121.2$^\circ$ and offset \obj\ to the upper right quadrant of the detector to keep
the tip-tilt star within the range available to the field steering
mirrors. 

The images were reduced following the standard procedures\cite{Fu12b}.
Before individual frames were finally combined, the camera distortion
was corrected using the on-sky distortion solution from observations of
the globular cluster
M~92\footnote{http://www2.keck.hawaii.edu/inst/nirc2/forReDoc/post\_observing/dewarp/}.
Five frames were rejected because of poor AO corrections, so the final
image combined 35$\times$80-s frames. It has a FWHM resolution of
0.18\arcsec, as measured from the stellar source located 10\arcsec\
W of \obj. The image was flux and astrometry calibrated
against the \K-band image from the VISTA Hemisphere Survey (VHS).
The \K-band image is shown in Figure~1. To highlight \obj,
we added contours to the Keck image after subtracting the foreground galaxies.

\obj\ was observed as part of our {\it Hubble} Space Telescope (\hst)
Cycle 19 SNAP Program (ID: 12488, PI: Negrello). Four images of 63~s 
were taken in the F110W
filter by the Wide Field Camera 3 (WFC3) on 2012 September 4 (UT). We
applied a four-point dithering pattern with a parallelogram primary
pattern shape, a point spacing of 0.572\arcsec, and a line spacing of 0.365\arcsec.
After passing the images through the standard WFC3 pipeline, we used the
Image Reduction and Analysis Facility (IRAF) {\sc MultiDrizzle} package on the four frames for distortion
correction, image registration, cosmic-ray rejection, and image
combination. A square interpolation kernel was used in the drizzling
process and the final reduced image has a pixel scale of 0.06\arcsec\ and 
a FWHM resolution of 0.24\arcsec.

Finally, we obtained seeing-limited \J\ and \K-band images of \obj\ with
the Long-slit Intermediate Resolution Infrared Spectrograph (LIRIS),
which is mounted on the Cassegrain focus of the 4.2-m William \herschel\
Telescope (WHT) (PI: Perez-Fournon). The observations took place on 2011 January 15 (UT). We
took 90$\times$20-s exposures in \K-band and 135$\times$40-s exposures
in \J-band. The images were reduced with the IRAF LIRIS data
reduction package {\sc
lirisdr}\footnote{http://www.iac.es/galeria/jap/lirisdr/LIRIS\_DATA\_REDUCTION.html}.
The FWHM resolutions are 0.9\arcsec\ and 0.8\arcsec\ for the final \J\
and \K-band images. The astrometry was calibrated against 2MASS. The
images were flux calibrated against the public VHS \J\ and \K-band
images of the same region. The WHT images are much deeper than the VHS images. 
The two components of \obj\ are easily separated
in the WHT images and are detected at 4$-$9$\sigma$.

Besides the morphology information from the high-resolution Keck and \hst\ images, all of these images were used to obtain photometry for both the foreground galaxies and \obj\ so that we can study 
their spectral energy distributions (SEDs).
Wherever detected, the two components of \obj\ show consistent configurations in different images. However, we do not detect the submillimeter and \co\ filaments between the two main components in these images. \objS\ appears much bluer in F110W$-$\K\ than \objN, which is a result of the ``clump'' that we identified at the southern end of \objS\ (Figure~\ref{fig:obs}). The clump is also detected in the archival Canada-France-Hawaii telescope optical images, and its SED is consistent with either a less obscured galaxy at $z = 2.3$ or a physically unrelated contaminating source. 

\subsection{Millimeter Photometry} \label{sec:mambo}

Millimeter continuum is useful to constrain Rayleigh-Jeans tail of the thermal dust emission. 
We observed \obj\ at 1.2\,mm with the Max-Planck Millimetre Bolometer
2 (MAMBO2\cite{Kreysa99}) at the IRAM 30\,m telescope on 2011 January 19 and 22
as part of the pool observations, achieving a 1$\sigma$ sensitivity of
$\sim$0.9\,mJy (PI: Perez-Fournon). \obj\ was clearly detected with a flux density of
$11.2\pm0.9$\,mJy. We note that some MAMBO2 observations taken around
the same time could have been affected by a ``drift'' signal of
$\sim$5\,mJy, but it is unlikely that the \obj\ observations were
influenced, because (1) we obtained consistent measurements from two scans that
were carried out three days apart and (2) the 1.2\,mm flux density is consistent with 
the best-fit modified blackbody from the photometry at other far-IR/millimeter 
wavelengths (see \S~\ref{sec:dust}). We also detected the continuum of \obj\ at 2.1~mm with a flux density of $1.2\pm0.2$~mJy as part of the \cofour\ observations with the PdBI (see \S~\ref{sec:pdbi}). 

\subsection{Archival X-ray-to-Radio Imaging} \label{sec:archival}

\obj\ was observed on 2006 July 4, 2006 July 22, 2008 July 30, and 2009
Jan 1 by \xmm\ as part of the Large Scale Structure
Survey\cite{Pierre04}. Two pointings (XMM-LSS\_42 and 45) in AO5 covered
\obj. We combined the MOS observations that contains \obj\ and obtained
a stacked image with a total integration time of 85~ks. It is
unfortunate that \obj\ happened to lie on the edges of both pointings 
where the sensitivity is low.
No source is detected at the position of \obj. In \S~\ref{sec:X-ray}, we
provide upper limits on the X-ray fluxes and luminosities.

To build the panchromatic SEDs of \obj\ and the foreground galaxies, we
obtained imaging data from the Galaxy Evolution
Explorer\cite{Morrissey07} (\galex, FUV, NUV), the Canada-France-Hawaii
Telescope Legacy Survey Wide\cite{Gwyn12} (CFHTLS-Wide, $u^*, g', r',
i'$, and $z'$), the VHS (\J, $H$, and \K), the \spitzer\ Wide-area Infrared
Extragalactic Survey\cite{Lonsdale03} (SWIRE; 3.6, 4.5, 5.6, 8.0, 24,
70, and 160~\um), the Wide-field Infrared Survey Explorer all-sky
survey\cite{Wright10} (\wise; 3.4, 4.6, 12 and 22~\um), and the FIRST survey\cite{Becker95}
(21~cm), in addition to our \hst\ (F110W), WHT (\J, and \K), Keck/NIRC2 (\K), \herschel\ PACS and SPIRE (70 to 500~\um), SMA (880~\um), MAMBO2 (1.2~mm), and IRAM/PdBI images (2~mm). In \S~\ref{sec:deblend}, we describe how we obtained the deblended photometry for \obj\ and the foreground galaxies.  

\begin{figure}[!t]
\begin{center} 
\includegraphics[width=8cm,clip]{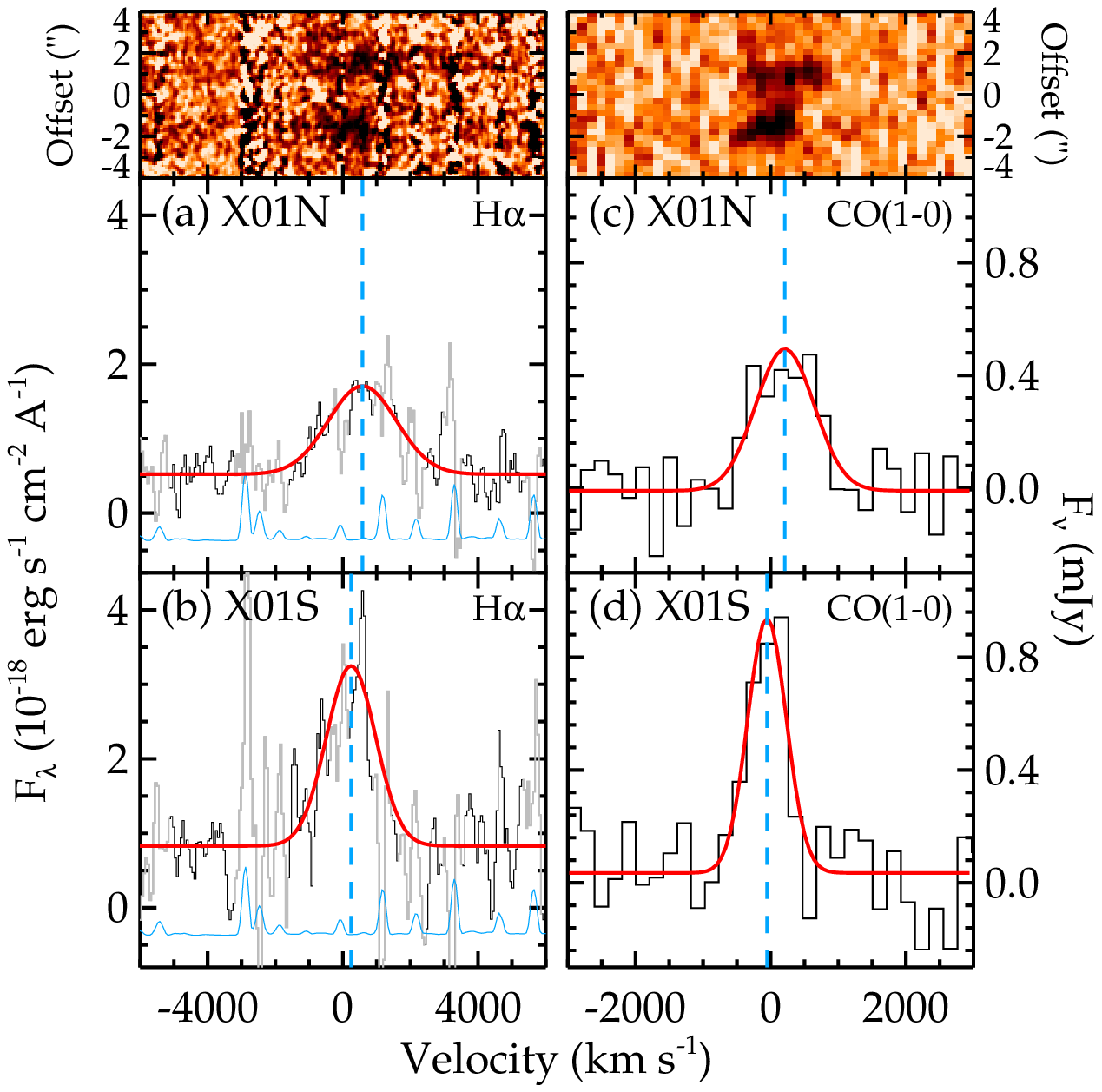}
\end{center} 
\caption{\textbf{ Keck/NIRSPEC near-IR and JVLA centimeter spectra of \obj}.
The velocities were computed against the systemic redshift
at $z = 2.308$. On the top, we use the 2D spectra to show the velocity offset 
between the two components of \obj. The vertical axis is along the spatial 
direction and NNE is up (slit PA = 24.3$^{\circ}$ and 22$^{\circ}$ for the H$\alpha$ and \co\ spectra, respectively). The \co\ 2D spectrum is extracted from the JVLA data cube using a 1.2\arcsec\ pseudo-slit centered on the two main components.  
The middle and bottom left panels show the H$\alpha$ spectra of \objN\ and \objS, along with
the best-fit Gaussians (red curves). Regions affected by strong sky
lines are plotted in light grey. A scaled sky spectrum is shown in blue.
The dashed vertical lines indicate the center of the best-fit Gaussian.
The right panels show the \co\ spectra along with the best-fit Gaussians
(red curves). \label{fig:spec}} 
\end{figure}

\begin{figure}[!t]
\begin{center} 
\includegraphics[width=8cm,clip]{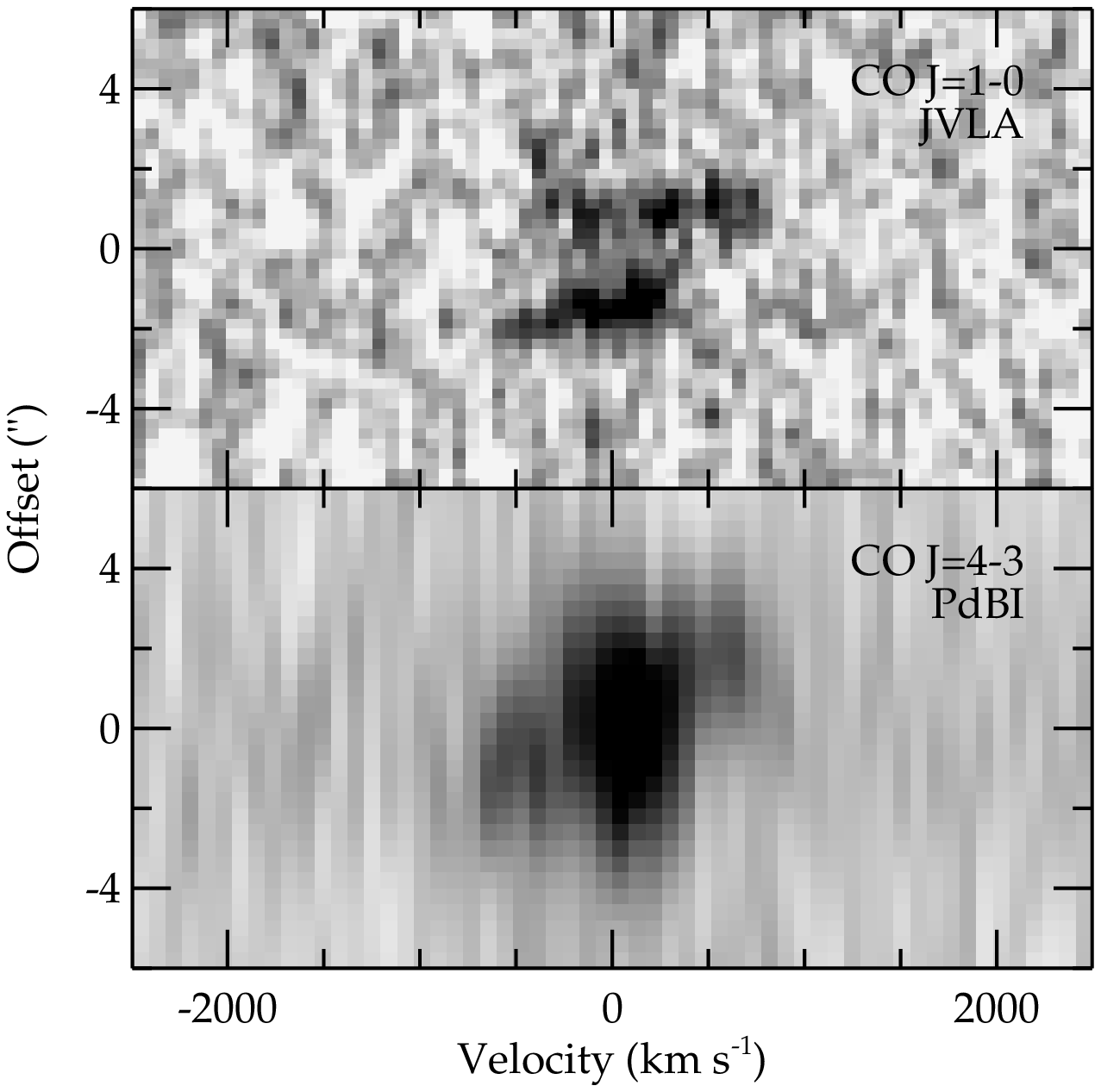}
\end{center} 
\caption{\textbf{JVLA and PdBI CO spectra of \obj}.
The velocities were computed against the systemic redshift
at $z = 2.308$. The \co\ 2D spectrum at the top is extracted from the JVLA data cube using a 1.2\arcsec\ pseudo-slit at PA = 22$^{\circ}$, while the \cofour\ spectrum at the bottom is extracted from the PdBI data cube using a 2.4\arcsec\ pseudo-slit at the same PA. Both panels clearly show the different velocity structures of the two components, strengthening the findings in Fig.~\ref{fig:spec}. \label{fig:spec2d}} 
\end{figure}

\subsection{Redshift from \cothree}

We used CARMA to search for \cothree\ emission towards \obj\
in the 3\,mm band\cite{Riechers11d} (PI: Riechers). The search was guided by 
the photometric redshift of $2.7\pm0.5$ (Dave Clements) obtained from
fitting the 250--1100\,\um\ SED. Observations were carried out under
good to excellent 3\,mm weather conditions during 4 tracks between
September 07 and 10 in 2010 in the D array configuration (6--127\,m baselines),
yielding an on-source (total) observing time of 7.8\,hr (13.3\,hr). Each
of the tracks targeted a different frequency setting, using the 3\,mm
receivers and the CARMA spectral line correlator with an effective
bandwidth of 3.7\,GHz per sideband (IF range:\ 1.2--4.9\,GHz) at
5.208\,MHz (15.6\,\kms\ at 100\,GHz) spectral resolution, yielding a
contiguous frequency coverage of 85--112\,GHz to search for redshifted
CO line emission. The nearby quasar J0239--025 was observed every
15\,minutes for complex gain calibration. Pointing was performed at
least every 2--4\,hr on nearby stars and radio quasars, using both
optical and radio modes. The bandpass shape was derived from
observations of the bright quasar 3C~84. Absolute fluxes were
bootstrapped relative to Uranus. The resulting calibration is estimated
to be accurate within $\sim$15\%.

We used the MIRIAD package for data reduction and analysis. The
combined data were imaged using "natural" baseline weighting, resulting
in a synthesized clean beam size of 5.6\arcsec$\times$4.2\arcsec\ (PA
149$^\circ$) and a rms noise level of 0.75\,mJy\,beam$^{-1}$ per
405.4\,MHz (1162\,\kms ) frequency bin.

A systematic inspection of the data cube both in the uv and image planes
yields a single, broad emission feature close to the phase center at the
position of the SMA continuum emission. We interpret this feature as 
the \cothree\ line emission ($\nu_{\rm
rest}$=345.7959899\,GHz, redshifted to $\nu_{\rm obs}=104.56\pm0.03$\,GHz)
towards \obj, yielding a redshift of $z =2.3073\pm0.0010$. We subsequently confirmed this redshift
by detecting \co\ line emission with the GBT and JVLA (see next
subsection) and \cofour\ line emission with the IRAM/PdBI
at the same redshift (see \S~\ref{sec:pdbi}).

\subsection{\co\ Spectroscopy} \label{sec:co10}

\begin{figure}[!t]
\begin{center} 
\includegraphics[width=8cm]{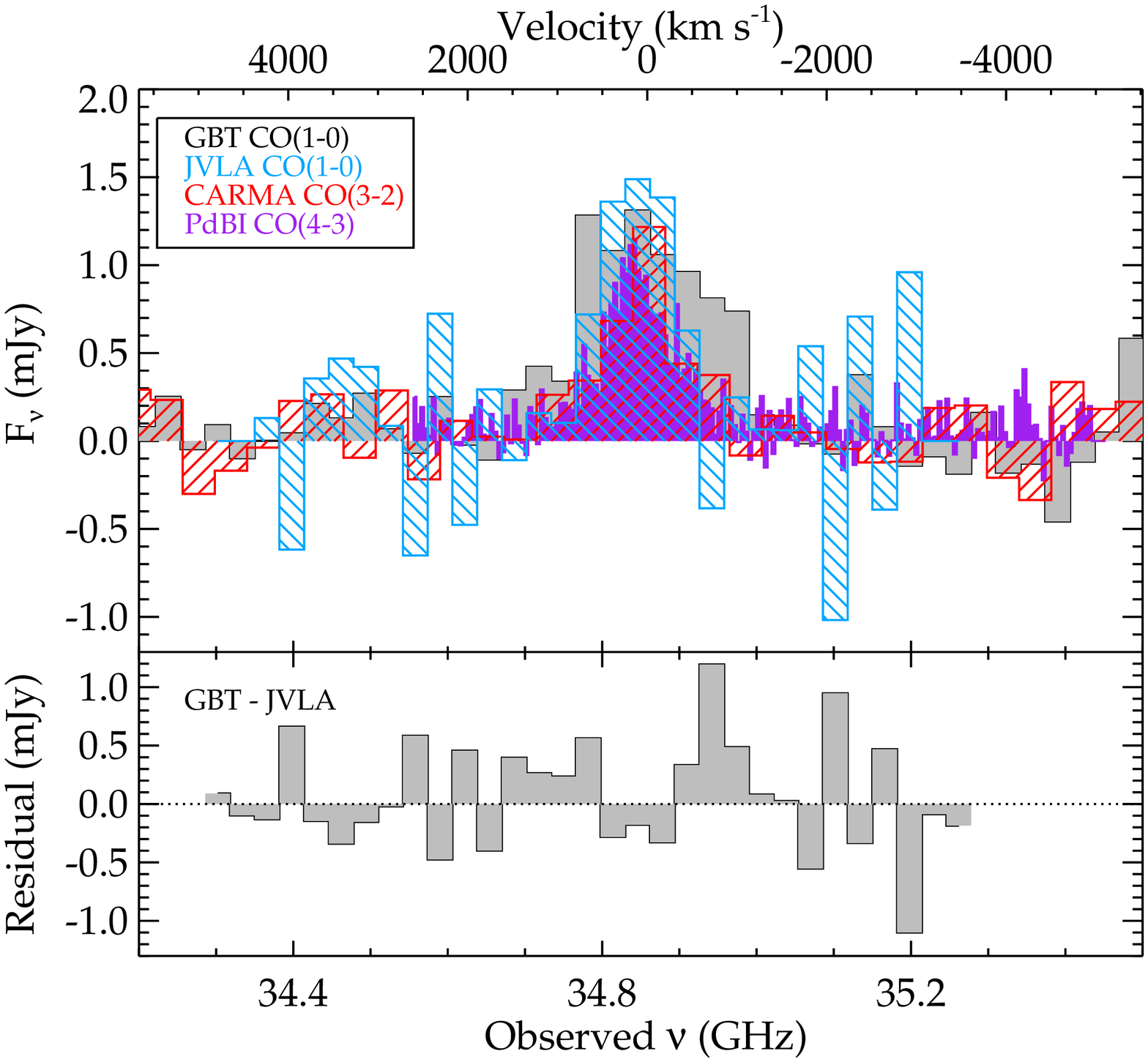}
\end{center} 
\caption{\textbf{ Comparison of CO line profiles}. The PdBI \cofour\ and CARMA \cothree\ 
lines have been scaled by $\nu_{1-0}/\nu_{J}$ in frequency and
$(\nu_{1-0}/\nu_{J})^2$ in flux density. The top axis shows the
velocity relative to the systemic redshift of $z = 2.308$. The JVLA, CARMA,
and PdBI spectra are extracted using 6\arcsec, 12\arcsec, and 15\arcsec-diameter apertures. \label{fig:colines}} 
\end{figure}

We obtained single-dish \co\ data at the 100\,m diameter GBT for
4.85\,hr on 2010 November 21 and 2.10\,hr on 2011 January 14 (Program
ID: GBT09C-70; PI: Harris). We used the Zpectrometer cross-correlation
spectrometer\cite{Harris10} attached to the facility Ka-band correlation
receiver. With instantaneous frequency coverage from 26.5 to 37.7\,GHz,
the spectrometer covers redshifts $2.1 \leq z \leq 3.5$ for the
115.27\,GHz \co\ transition. Over the spectrometer's band the
FWHM beam size ranged from 27\arcsec\ to 16\arcsec, and the correlator's
20\,MHz bandwidth corresponded to a spectral resolution of 234 to
157\,\kms. Cross-correlation produces spectra that are the difference in
power between the receiver's two beams on the sky, which are separated
by 78\arcsec.  The correlation receiver architecture,
electrical phase switching, and optically switching the source between
the receiver inputs with sub-reflector motion on a 10\,sec cycle greatly
reduces electronic and atmospheric fluctuations. We switched the telescope
between two targets close on the sky on an 8\,min cycle to compensate for
optical imbalance between the two beams. 
We established the absolute flux scale with hourly pointing on quasar
0108+0135; spectra of this source before and after pointing also
provided average corrections for pointing drift and atmospheric
transmission. The flux density of 0108+0135 was 1.8\,Jy during the
observations, calibrated against 3C~48, whose flux density is 0.74\,Jy
at 34.8~GHz, following the Astronomical Almanac (2011). 
The \co\ line from \obj\ is clearly detected at $34.85\pm0.01$~GHz,
corresponding to $z = 2.3074\pm0.0008$. The FWHM beam size of the GBT is
22\arcsec\ at this frequency.
 
We also obtained high-resolution \co\ data with the JVLA (PI: Ivison). Observations
were carried out dynamically during good weather conditions in 2012
January (DnC configuration), March (C configuration), and June (B
configuration). The bright compact calibration source, J0241$-$0815, was
observed every few minutes to determine accurate complex gain solutions
and bandpass corrections. 3C~48 was also observed to set the absolute
flux scale, and the pointing accuracy was checked locally every hour.
At the redshift of \obj, the receivers gave 1,024\,MHz coverage with 512
$\times$ 2\,MHz channels. The data were reduced using {\sc
AIPS}\cite{Ivison11}. Because the {\sc AIPS/SETJY} task adopts a higher
flux for 3C~48 ($S$ = 0.83~Jy at 34.8~GHz; i.e., 12\% higher than that
assumed for the GBT/Zpectrometer data), we scaled the GBT flux densities
by the ratio of the two adopted 3C~48 flux densities (0.83~Jy/0.74~Jy) 
for a fair comparison.
In the end, the channels were imaged over a $512\times 512\times$
0.1\arcsec\ field, with natural weighting ({\sc robust = 5}), to form a
$512^3$ cube centered on \obj. Integrating over the 80 channels ($\sim 1,400$\,\kms) that
contain line emission yielded an r.m.s.\ noise level of 25\,\uJy\,beam$^{-1}$. 

The {\sc clean}ed and velocity-integrated \co\ map is shown in
Figure~\ref{fig:obs} as blue contours. The beam is 0.83\arcsec$\times$0.77\arcsec\ at
PA = $72^{\circ}$. Similarly to the SMA data, \obj\ was resolved into
two main components separated by $\sim$2.8\arcsec. The northern component shows an extension of 
low-surface-brightness filament towards the southern component. This filament has a line flux of $S_{\rm CO}\Delta V = 0.20\pm0.04$~\,Jy\,\kms, which is 2.5 times lower than that of the northern component. 
We extracted the CO spectra with 1.2\arcsec\ and 2.0\arcsec-diameter apertures centered on
the northern and southern components, respectively. Based on the
best-fit Gaussian models of the velocity-integrated map, we then applied
aperture corrections of 1.7$\times$ and 1.3$\times$ to the spectra of
the northern and southern components, respectively. The two \co\ components have
significantly different redshifts and line profiles (Figure~\ref{fig:spec} and Table~\ref{tab:prop}),
indicating that they are two galaxies that are merging, instead of lensed images from a single 
source. This result is confirmed by the Keck H$\alpha$ spectra (\S~\ref{sec:ha}; see Figure~\ref{fig:spec}) and the PdBI \cofour\ data cube (\S~\ref{sec:pdbi}; see Figure~\ref{fig:spec2d}).  

The JVLA spectrum integrated over a 6\arcsec-diameter aperture gives a
line width of $\Delta V_{\rm FWHM}=840\pm160$\,\kms\ and a line flux of
$S_{\rm CO}\Delta V = 1.7\pm0.3$\,Jy\,\kms\ at $z = 2.3079\pm0.0007$. 
In comparison, the \co\ measurements from the GBT are: $\Delta V_{\rm FWHM}=1670\pm140$\,\kms\ and $S_{\rm CO}\Delta V = 2.3\pm0.3$\,Jy\,\kms\ at $z = 2.3074\pm0.0008$. 
As shown in Figure~\ref{fig:colines}, the discrepancy between JVLA and GBT \co\ line intensity is insignificant ($0.6\pm0.4$\,Jy\,\kms), and it mostly occurs in the wings, where the S/N is low. 
While some flux from the extended wings could have been resolved out in the high-resolution JVLA data, we primarily adopt the JVLA measurements when deriving the properties of \obj\ to be conservative about its gas mass.

\subsection{\cofour\ Spectroscopy and 2~mm continuum} \label{sec:pdbi}

We observed \obj\ with the PdBI tuned at 139.414 GHz, as part of the millimeter follow-up of the HerMES gravitational lens candidates (PI: Cox). The 
Widex correlator bandwidth (3.6 GHz) allows covering the \cofour\ emission line and the underlying 2~mm continuum. The observations were carried out in September 2010 in the compact D array configuration with 4 antennae, with a total on-source integration of 3.7~hours. System temperatures during the observations were in the range 70$-$140~K, and precipitable water vapor around 2~mm. Data analysis was carried out with GILDAS. Maps were obtained using natural weighting, resulting in a synthesized beam of 8.9\arcsec$\times$3.1\arcsec\ and PA = 140 deg. The noise level is 0.8~mJy~beam$^{-1}$ in 20~MHz channels.

We detected a bright \cofour\ line peaking at a redshift $z=2.3081\pm0.0002$, consistent with the redshift obtained from the CARMA, the JVLA, and the GBT observations. The line is detected at a high significance (30$\sigma$; Figure~\ref{fig:colines}). The spatially resolved velocity structure is consistent with that observed in the JVLA data cube (Figure~\ref{fig:spec2d}), further confirming that \obj\ is a merger. 

In addition to the \cofour\ line, we also detected the continuum at 2.1~mm with a flux density of $1.2\pm0.2$~mJy.

\subsection{K-band H$\alpha$ Spectroscopy} \label{sec:ha}

We obtained a $K$-band spectrum of \obj\ on 2010 November 21 (UT) with
the Near Infrared Spectrometer (NIRSPEC\cite{McLean98}) on the Keck\,II telescope (PI: Casey).
We used the low-resolution mode with NIRSPEC-7 ($K'$) filter and a cross-dispersion angle of 35.49$^{\circ}$, which gave a wavelength range of
1.96$-$2.40\,\um. The 0.76\arcsec\ ($R \sim 1450$) slit was oriented at
a PA of 24.3$^{\circ}$ east of north to capture both components of \obj.
We took eight frames of 600\,s each (10 coadds of 60\,s exposures),
adopting the standard ``ABBA'' dithering sequence with a step of
15\arcsec\ along the slit. HIP\,34499 (\K\ = 10.6 AB) was observed using
the same setup for telluric correction. 

We used the REDSPEC IDL package\footnote{http://www2.keck.hawaii.edu/inst/nirspec/redspec.html} to perform spectral and spatial rectification on all of
the frames. Most of the sky background and fringing patterns were
removed by subtracting the two nodding positions, and the residual sky
background was removed by shifting the rectified and median-combined
A$-$B 2D spectrum along the spatial direction by 15\arcsec\ and subtract it 
from the unshifted 2D spectrum.
One-dimensional spectra were extracted with 1.8\arcsec-wide apertures.
We detected a broad H$\alpha$ line as well as the continuum in both
\objN\ and \objS. We calibrated the flux of the spectra with the \K-band
photometry from the WHT \K-band image
(\S~\ref{sec:nir}). In total, the H$\alpha$ lines
contribute to $\sim$12\% ($2.7\pm0.4$\,\uJy) of the \K-band flux of
\obj. Consistent with the results from the CO lines, the H$\alpha$ spectra also show different redshifts and line profiles for \objN\ and \objS\ (Fig.~\ref{fig:spec}). Table~\ref{tab:prop} summarizes the measurements from the best-fit Gaussians. 

\subsection{Optical Spectroscopy of Foreground Galaxies} \label{sec:lris}

We obtained optical spectra of the two foreground lensing galaxies (G1
and G2 in Figure~\ref{fig:obs}) on 2011 March 1 (UT) with the Low
Resolution Imaging Spectrometer\cite{Oke95} on the Keck\,I telescope (PI: Bridge). We
took two 540\,s exposures through a 1.5\arcsec\ slit at PA =
275.2$^{\circ}$ (i.e., aligned with G1 and G2). On the blue side, we
used the 600 groove~mm$^{-1}$ grism blazed at 4000~\AA; while on the red
side, we used the 400 groove~mm$^{-1}$ grating blazed at 8500~\AA\
tilted to a central wavelength of 7779~\AA. The 560~nm dichroic was
used. The wavelength coverages are 3000$-$5600~\AA\ and 5400$-$10200~\AA,
and the spectral resolutions are 6 and 10~\AA\ (FWHM) for the blue and red,
respectively. The spectra were taken at an airmass of $\sim$1.75, and
the atmospheric seeing was $\sim$0.6\arcsec\ at 0.5~\um. 

The data were reduced and extracted with IRAF. From the [O\,{\sc
iii}]\,$\lambda\lambda$4959,5007 lines of G1, and H$\alpha$ and [N\,{\sc
ii}]$\lambda\lambda$6548,6583 lines of G2, we determined their redshifts
of $z_{\rm G1} = 0.6546\pm0.0001$ and $z_{\rm G2} = 0.5020\pm0.0001$, respectively.

\section{Deblending the Photometry} \label{sec:deblend}

\begin{figure*}[!t]
\begin{center} 
\includegraphics[width=12cm]{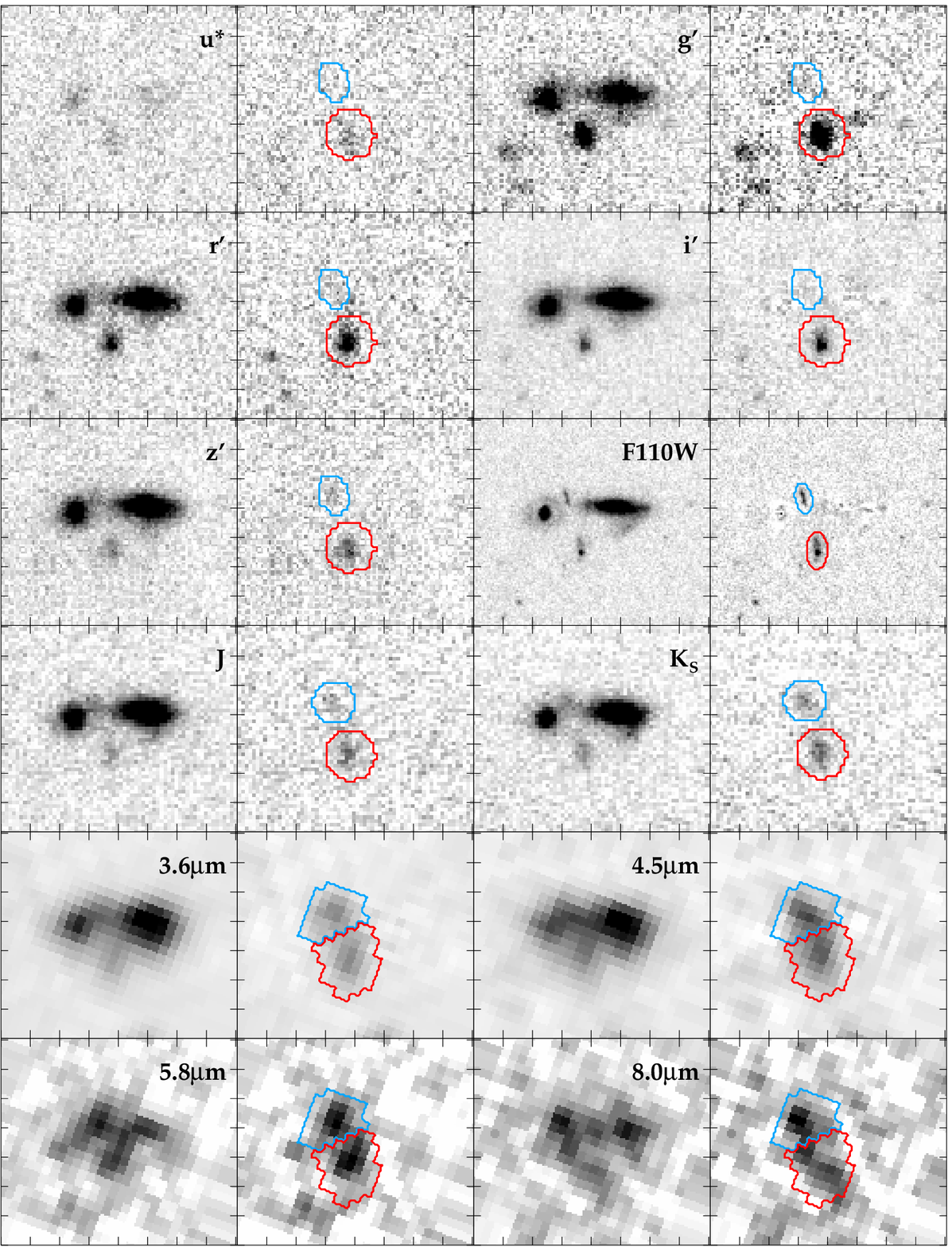}
\end{center}
\caption{\textbf{ Deblending the photometry from the $u*$-band to 8\,\um}. All
images are 16\arcsec$\times$14\arcsec\ and are aligned in astrometry so
that N is up and E is to the left. The tickmarks are spaced at intervals of
2\arcsec. For each filter, the original image is shown on the left and
the residual image after subtracting the two foreground galaxies is shown on the right.  
The apertures used for our
photometry are outlined in the residual images (blue $-$ \objN; red $-$
\objS).
\label{fig:galfit}} 
\end{figure*}

Because of the proximity of \obj\ to the foreground galaxies on the sky, we need to deblend the various
sources in some of the images to obtain robust photometry.
Between the $u*$-band and IRAC 8.0~\um, we model the various
components in individual images with \galfit\cite{Peng10}. We use high-S/N PSF stars near 
\obj\ that are unsaturated and relatively isolated. For images between the $u*$-band and the \K-band, we use a star 10\arcsec\ south-southwest of \obj. For the IRAC images, we use a star 7.7\arcmin\ north of \obj. We model the foreground galaxies as \sersic\ profiles. The foreground galaxy G2 shows more complex structures (e.g., the tail to the SW) than G1, so we include multiple \sersic\ profiles with their relative positions fixed to those determined from the high-resolution \hst\ image. Instead of masking out \obj, we include multiple Gaussians to approximate its clumpy morphology and fit them simultaneously with the foreground galaxies. We obtain reasonably good fit with $\chi^2/{\rm dof} \sim 1.0$ for all of the images except for the \hst-F110W image ($\chi^2/{\rm dof} \sim 1.4$). Although the \hst\ image is the most challenging to model because of the detailed structures of the foreground galaxies, the photometry is very robust because the components do not overlap as much as the seeing-limited images and IRAC images.  

The total fluxes of G1 and G2 are obtained from the best-fit models. 
We then subtract the models of
G1 and G2 from the original images to reveal emission from \obj\
(Figure~\ref{fig:galfit}). Finally, we measure the fluxes from the
residual images with two apertures placed around the northern and
southern components. As shown in Figure~\ref{fig:galfit}, the apertures 
are large enough to enclose all of the detectable flux, yet are small enough 
to avoid including too much background noise. It is clear that \objS\ is much brighter
than \objN\ in the optical bands but the two are similarly bright in the
near-IR, suggesting either less dust obscuration in \objS\ or that the
southern clump of \objS\ is not part of the system (Figure~\ref{fig:obs}).
Since we cannot separate the two parts of \objS\ at wavelengths other than
the \hst/F160W and Keck/\K-band, we decided to include both in the
photometry. Future \hst\ imaging and spectroscopy are needed to reveal
the nature of the southern clump of \objS.

\galex\ barely detected one source at the position of \obj\ in the NUV band. Considering 
the spatial resolution in the NUV band is 6\arcsec\ in FWHM, G1, G2, and \obj\ 
are blended together. The source
has a total NUV flux density of $0.67\pm0.18$\,\uJy\ and is undetected at FUV.
We can safely assume that the contribution of \obj\ is negligible
because both filters are shorter than the Lyman break at $z = 2.31$. Unable to
deblend the flux between G1 and G2, we assume the 3$\sigma$ upper limit
(0.54\,\uJy) at NUV for all of the components. We do not use the FUV
upper limits because of its limited constraints on the SED. 

Deblending is unreliable for the \wise\ images because of the low S/N
and resolution. The best-fit SED models of G1 and G2 (see
\S~\ref{sec:sedlens}) suggest 
that most of the 11\,\um\ emission and
$\sim$25\% of the 22\,\um\ could arise from the polycyclic aromatic
hydrocarbons (PAHs) and dust continuum of the foreground galaxies,
although the models have large uncertainties due to the lack of
constraints at long wavelengths. Considering the large uncertainties
involved in estimating the foreground contamination at 11\,\um\ and that
the other three \wise\ channels overlap and are consistent with higher
quality \spitzer\ data, we opt not to use the \wise\ data for the SED
modeling. 

At wavelengths longer than $\sim$20\,\um, deblending is impossible
because of the limited spatial resolution of the available data, except
at 880\,\um\ where we have the high resolution SMA data. However, we 
do not need to subtract the foreground galaxies here,
because \obj\ should be dominating the total flux as the SEDs of the
foreground galaxies rapidly decline at these wavelengths. Indeed, the foreground galaxies are
not detected in our SMA image. \obj\ is resolved in the SMA image
(Figure~\ref{fig:obs}), so we can directly measure the 880\,\um\ fluxes of
\objN\ and \objS. Without high-resolution data at
other wavelengths, we assume the fluxes can be divided in the same
proportion among the three components as at 880\,\um\ --- 34\% in
\objN, 31\% in \objS, and 35\% in the filaments (see \S~\ref{sec:sma}). 
Clearly, we have assumed that the dust temperatures from the three components 
are the same, which may not be true in reality. 

We opt not to use the SWIRE 160~\um\ data because \obj\ is severely blended with another brighter source to the NW. And due to the complex structures in the background of that image, we cannot extract a reliable flux for \obj. Instead, we use the PACS 160~\um\ data that is much deeper and has much better spatial resolution.

We did not obtain useful measurements of \obj\ in the VHS $H$-band because of image artifacts and at MIPS 160\,\um\ because of blending and complex structures  in the background. 
\obj, G1, and G2 are undetected at FUV, NUV, MIPS 70\,\um, and 21~cm. The MIPS 70\,\um\ upper limit is consistent with the 2$\sigma$ detection made with the PACS 70\,\um\ image.  

We list the deblended photometry in Tables~\ref{tab:sedg1g2} and \ref{tab:sedobj}. In the photometric errors, we have included the following
flux calibration uncertainties: 3\% for CFHTLS\cite{Gwyn12}, 2\% for
VISTA and WHT\cite{Cohen03}, 3\% for IRAC\cite{Reach05}, 4\% for MIPS
24\,\um\cite{Engelbracht07}, 5\% for MIPS 70\,\um\cite{Gordon07}, 5\% for PACS 70, 110 and 160~\um, 
7\% for SPIRE\cite{Swinyard10} (confusion noises are also included in the total errors for PACS and SPIRE\cite{Nguyen10}), 10\% for SMA, and 15\% for MAMBO2 and PdBI.

\section{Lens Modeling} \label{sec:lensing}

The observed configuration and morphology of \obj\ do not provide constraints 
on the lensing potential, except that we can rule out models that produce 
multiple images of a single source (i.e., the strong lensing cases).
To constrain the lensing magnifications that the foreground galaxies
impose on \obj, we need to estimate the dark matter halo masses of G1 and G2 from
their stellar masses. The standard method to measure stellar masses is
to fit the observed UV-to-near-IR SEDs with stellar population
synthesis models.  

\subsection{Stellar and Halo Masses of the Lensing Galaxies} \label{sec:sedlens}

\begin{figure}[!t]
\begin{center} 
\includegraphics[width=8.5cm]{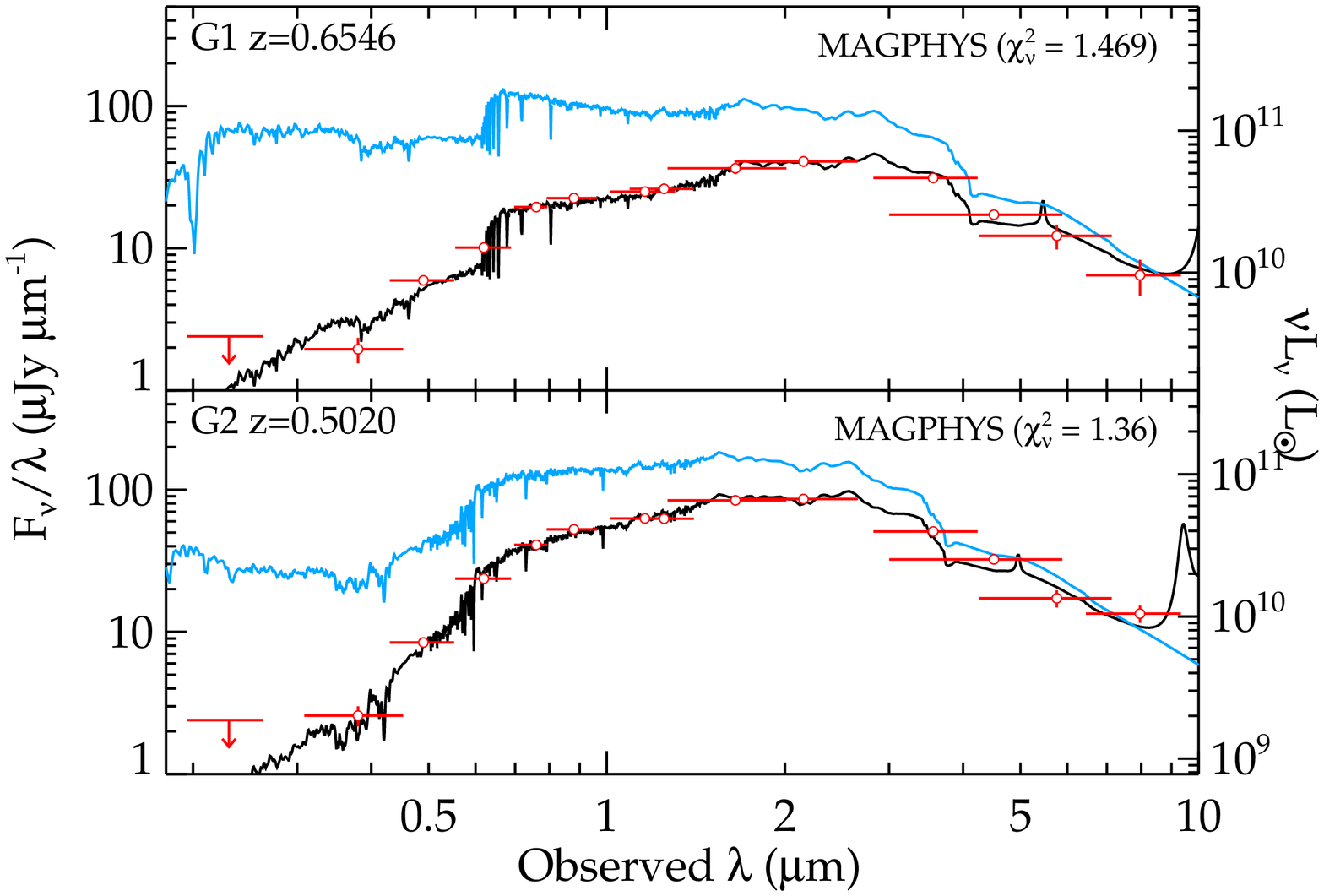}
\end{center} 
\caption{\textbf{ SEDs of the foreground galaxies G1 (top) and G2 (bottom)}.
The best-fit stellar population synthesis models are plotted as black
solid curves, and the corresponding intrinsic models without dust
extinction are plotted in blue. Labeled are the spectroscopic redshifts
that we obtained from the Keck/LRIS and the reduced $\chi^2$ values of
the best-fit models.  \label{fig:sedlens}}
\end{figure}

We use the \magphys\ software\cite{Cunha08} to fit evolutionary
population synthesis models to the SEDs. We used the default 25,000
model spectra computed using the 2007 version of the Bruzual and Charlot 
stellar population synthesis code\cite{Bruzual03}. The model spectra were computed at ages
between 0.1 and 13.5~Gyr for various metallicities and dust absorption
optical depths, assuming a Chabrier IMF\cite{Chabrier03} and
exponentially declining or constant star formation histories with random
bursts. Models with ages greater than the age of the Universe at the redshift of
the object are excluded. The software treats dust attenuation of young
and old stars separately, following a two-component
model\cite{Charlot00} assuming that stars younger than
$10^7$\,yr are embedded in their birth clouds so these stars are attenuated 
by dust in both the birth cloud and the interstellar medium (ISM). It then consistently computes
the re-radiated mid-to-far-IR dust emission in stellar birth clouds and
the ambient diffuse ISM, based on the power of the attenuated stellar
emission at shorter wavelengths. A $\chi^2$ and the corresponding
probability exp($-\chi^2/2$) are computed for each model. Finally, a
marginalized likelihood distribution is built for each parameter. We
adopt the best-fit parameter as the median of the likelihood
distribution, which is usually close to the value corresponding to the
minimum $\chi^2$, and its $\pm$1$\sigma$ confidence interval as the
16th$-$84th percentile range. 

We fix the galaxies to their spectroscopic redshifts
from the Keck/LRIS spectra (\S~\ref{sec:lris}; $z_{\rm G1} = 0.6546$ and
$z_{\rm G2} = 0.5020$). Both G1 and G2 are undetected in the SMA
880\,\um\ image, so we use the 3$\sigma$ upper limit at 880\,\um\ to
put some constraints on the dust emission, and therefore, on the amount
of dust extinction.

Figure~\ref{fig:sedlens} shows the best-fit models. Using the 16th,
50th, and 84th percentiles of the marginalized likelihood distributions,
we estimated stellar masses of log(\ms/\msun) = $10.72\pm0.06$
and $10.86\pm0.10$ for G1 and G2,
respectively. The stellar masses correspond to halo masses of
log($M_{\rm halo,G1}$/\msun) = $12.56\pm0.26$ and log($M_{\rm
halo,G2}$/\msun) = $12.81\pm0.33$, adopting the \ms$-$\mh\ relation at
their redshifts from abundance matching\cite{Behroozi10} and including 
the intrinsic scatter of the relation.

\subsection{Magnification Factors} \label{sec:magnification}

\begin{figure}[!t]
\begin{center} 
\includegraphics[width=7cm]{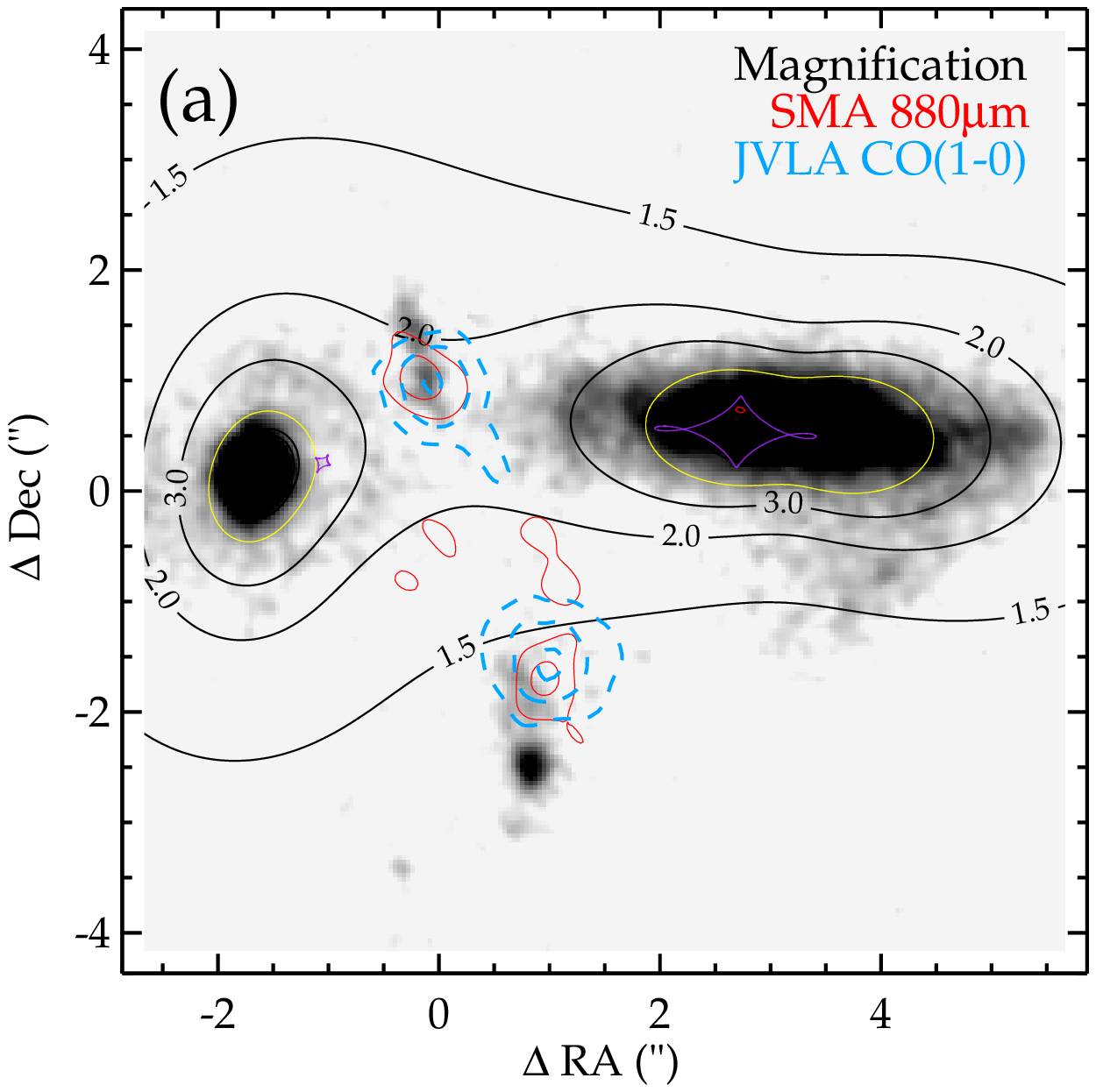}
\includegraphics[width=7cm]{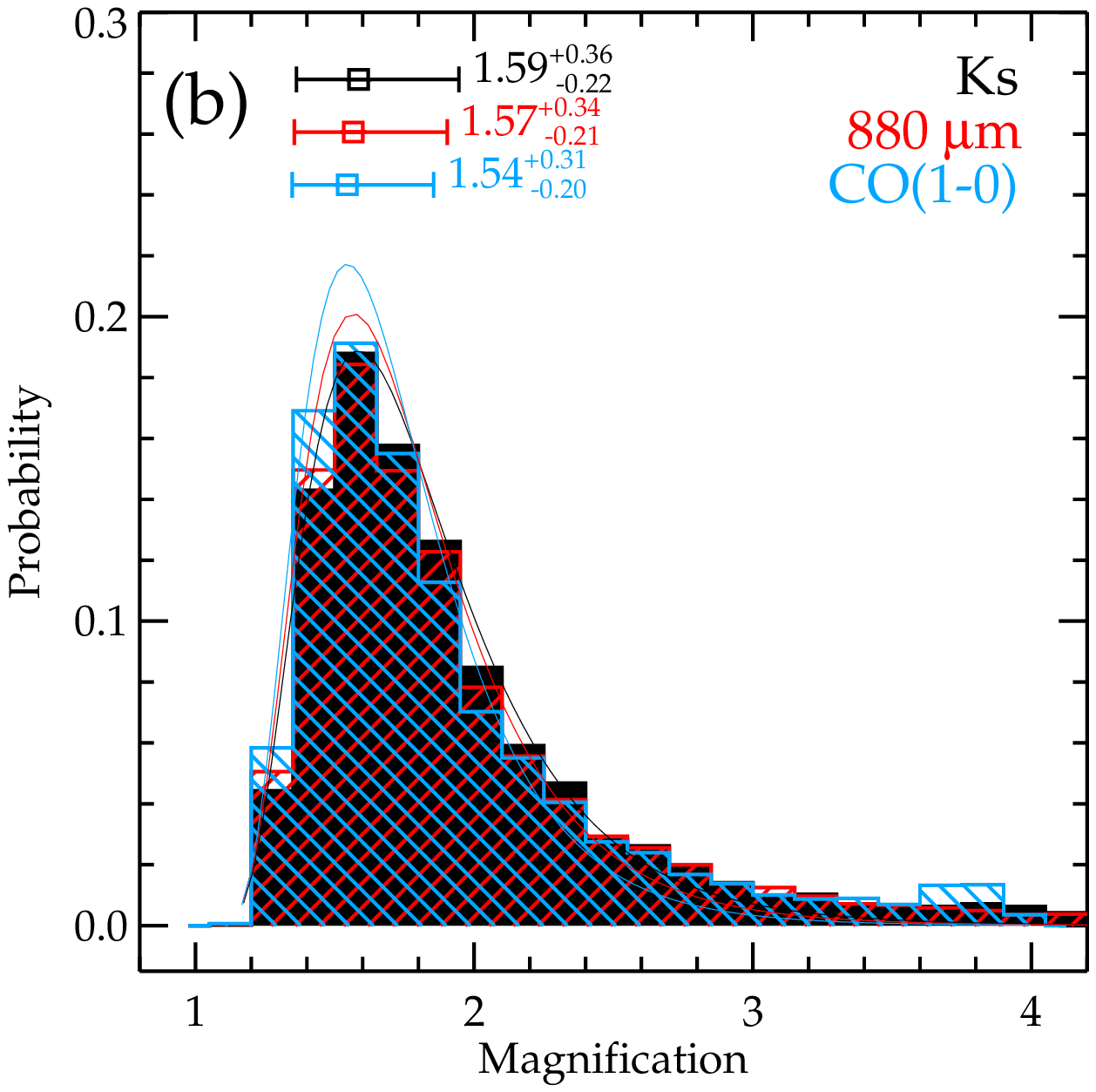}
\end{center} 
\caption{\textbf{ Magnification map and magnification factors}. (a) Similar
to Figure~\ref{fig:obs}, but here we overlay the image-plane magnification
map as black contours on the \hst/WFC3 F110W image. The magnification
map is computed assuming the median halo masses and SIE profiles.
Critical curves are in yellow and caustics are in purple. (b)
Distributions of the total magnification factors for the \K-band
(black), the dust continuum at 880\,\um\ (red), and the molecular gas as
probed by \co\ (blue). On the top, we show their $\pm$1$\sigma$ confidence
intervals along with the median magnification (boxes with error bars)
from log-normal fits (solid curves).  \label{fig:mag}} 
\end{figure}

\begin{figure}[!t]
\begin{center} 
\includegraphics[width=9.5cm]{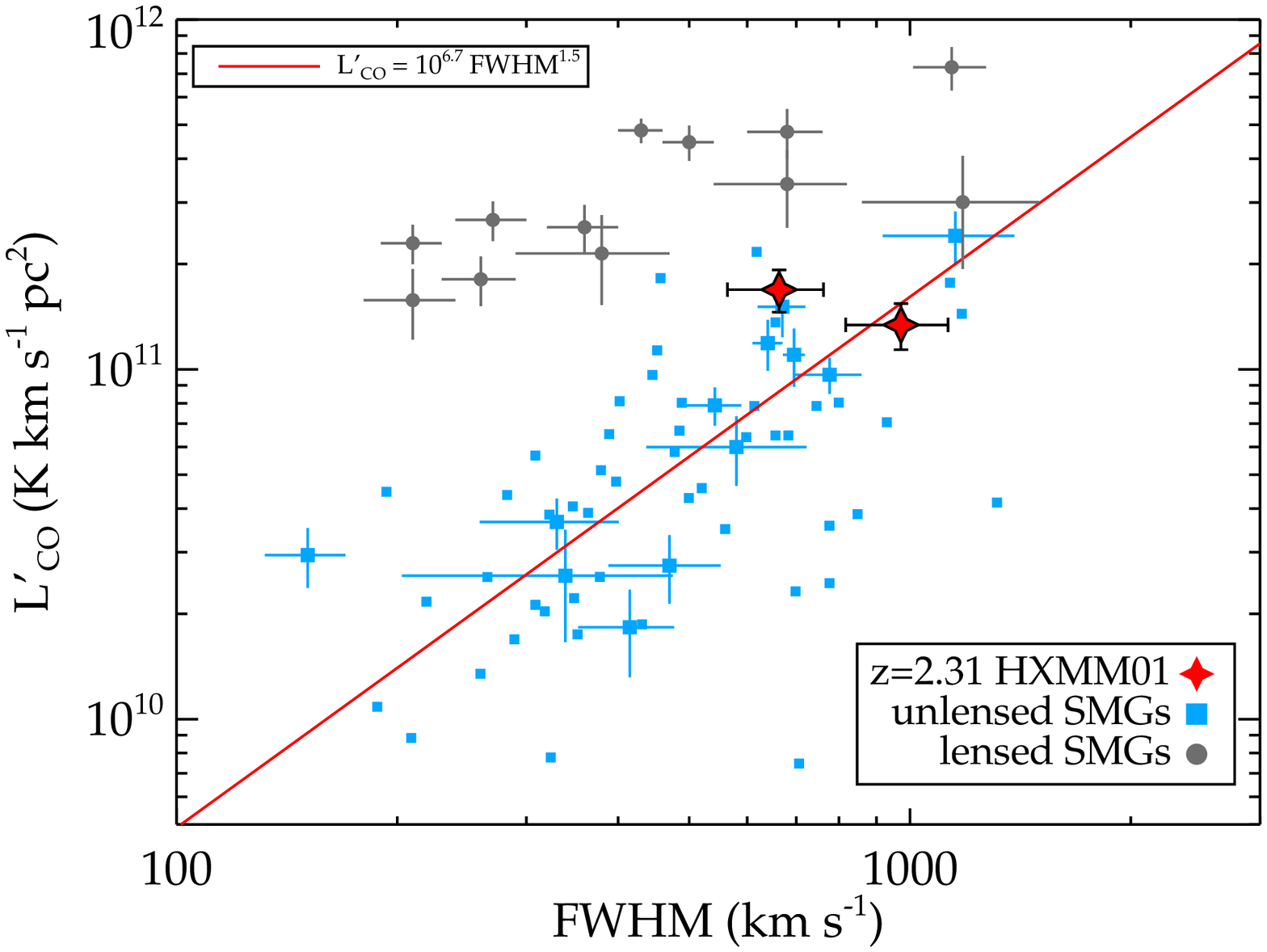}
\end{center} 
\caption{\textbf{$L'_{\rm CO}$ vs. FWHM for \co\ of lensed and unlensed SMGs}. The two red stars show the measurements of \objN\ and \objS\ from the JVLA data cube, which have not been corrected for lensing. The big blue squares with error bars are unlensed and lensing-corrected SMGs with \co\ measurements\cite{Harris10,Carilli10,Ivison11,Riechers11b}, and the small blue squares are mostly SMGs with higher $J$ CO line measurements converted to \co\ using mean observed ratios\cite{Carilli13}. The red line shows the best-fit relation for unlensed SMGs. The grey filled circles with error bars are the GBT \co\ measurements\cite{Harris12} of the brightest lensed SMGs in the H-ATLAS survey. 
The two main components of \obj\ lie on the same relation as unlensed SMGs, suggesting that they are mildly ($\mu\lesssim2$) magnified, consistent with our lens modeling results (\S~\ref{sec:magnification}). 
\label{fig:lco_fwhm}} 
\end{figure}

We constrain the lensing magnification factors ($\mu$) in a Monte-Carlo fashion.
We start by generating a distribution of 5000 pairs of stellar masses for G1 and G2
that follow the posterior likelihood distributions from \magphys. We
assign dark matter halo masses using the \ms$-$\mh\ relations from
abundance matching\cite{Behroozi10}. The 0.15\,dex intrinsic scatter in
the \ms$-$\mh\ relation at fixed halo mass is also included. We adopt
the singular isothermal ellipsoid (SIE) density profile for the lenses
and assume them to follow the centroid, ellipticity, and PA of the
light distribution, as supported by
observations\cite{Koopmans06,Sluse12}. For each pair of virial masses
(baryonic$+$dark matter), we compute the image plane magnification map
for sources at $z = 2.31$ with {\sc lenstool}
(Figure~\ref{fig:mag}$a$)\cite{Jullo07}. We then compute the luminosity
weighted magnification factors for \obj, \objN, and \objS\ using the
PSF/beam-deconvolved light-distribution models. We compute $\mu$ for the \K-band,
880\,\um, and \co\ images separately because there are significant spatial
offsets in the sources across these wavelengths. After excluding the strong lensing
cases where models generate more images then observed, the distributions
of magnification factors and source-plane projected separations are fit
with log-normal functions to find the medians and the 1$\sigma$
uncertainties (Figure~\ref{fig:mag}$b$). 

We find that the total luminosity-weighted magnification of \obj\ is only about 1.6. The
northern component (\objN) is more magnified ($\mu \sim 1.8$) than the
southern component (\objS; $\mu \sim 1.4$). The source-plane separation
of the two components in \K-band is just 2.2\arcsec\ (19~kpc), which is
35\% smaller than the image plane separation (3.4\arcsec\ = 28~kpc).
Table~\ref{tab:prop} lists the results. We find similar results using
the NFW profile\cite{Navarro96}, although the estimated magnifications
are $\sim$20\% lower than using the SIE profile. Despite the observed
spatial offsets among the gas, dust, and stellar components, their
magnification factors agree within 10\%. Therefore, we can safely ignore
differential magnification in this source.

As shown in Figure~\ref{fig:lco_fwhm}, the \co\ luminosities and line widths of \obj\ also suggest only mild ($\mu \sim 1.8$) magnifications from the foreground galaxies, because their \co\ luminosities are similar to those of unlensed SMGs with the same line widths. 

\section{Physical Properties of \obj} \label{sec:properties}

In this section we determine the physical properties of \obj\ from the
SEDs, the high-resolution CO and dust continuum images, the CO spectra,
and the X-ray images. Because the lensing magnifications are small ($\mu
< 2$) and have large uncertainties, we list the lens-amplified
properties in units of \mui\ instead of correcting for $\mu$. Our
derived physical parameters are summarized in Table~\ref{tab:prop} along
with the estimated magnification factors, so that readers can easily correct
for $\mu$ to obtain the intrinsic properties. Note that some parameters
are unaffected by lensing --- e.g., surface densities, temperatures, and
mass or luminosity ratios --- so magnification correction is
unnecessary. 

\subsection{Stellar Mass and Star Formation Rate} \label{sec:sedobj}

\begin{figure*}[!t]
\includegraphics[width=6.2cm]{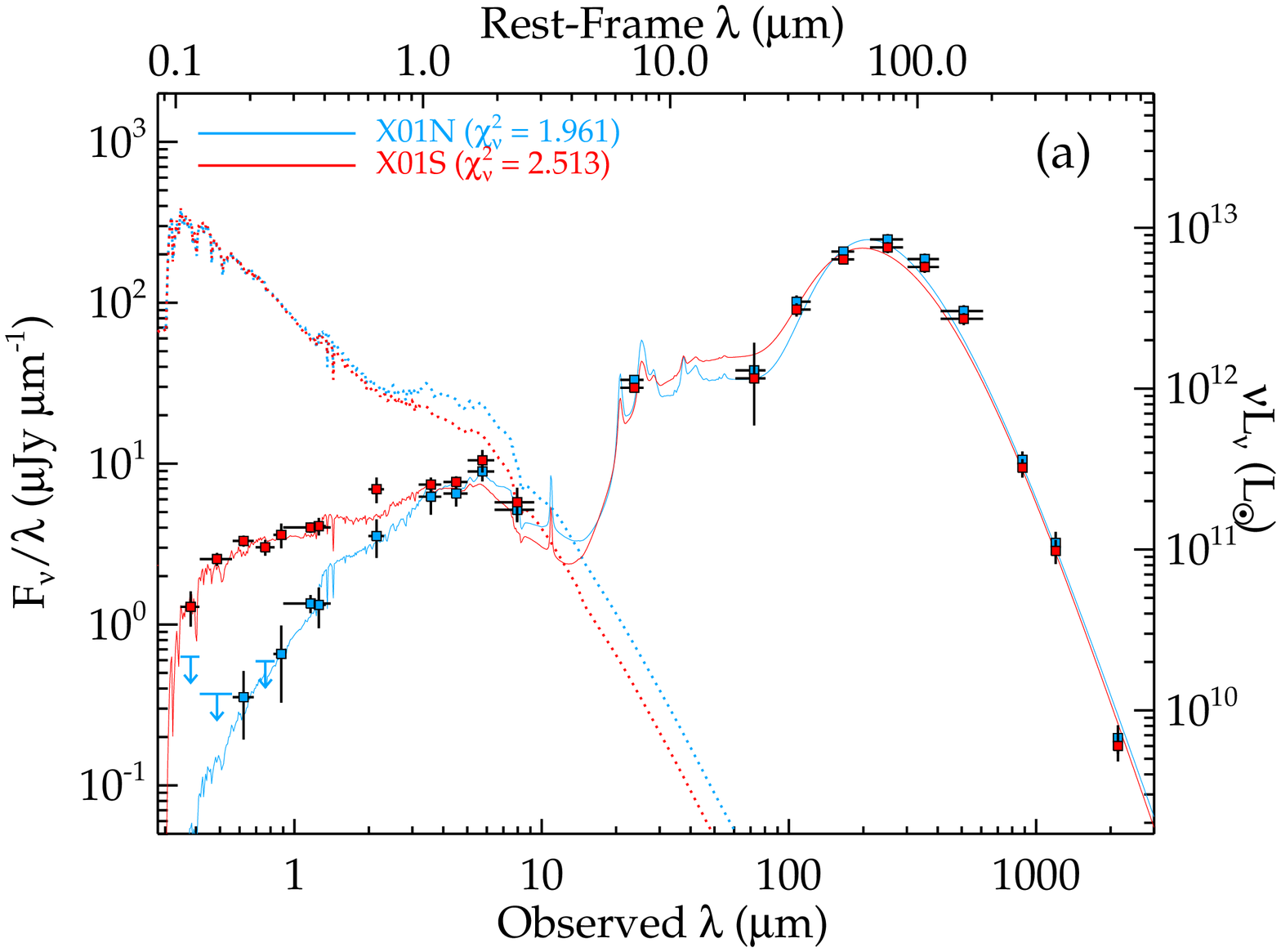}
\includegraphics[width=6.2cm]{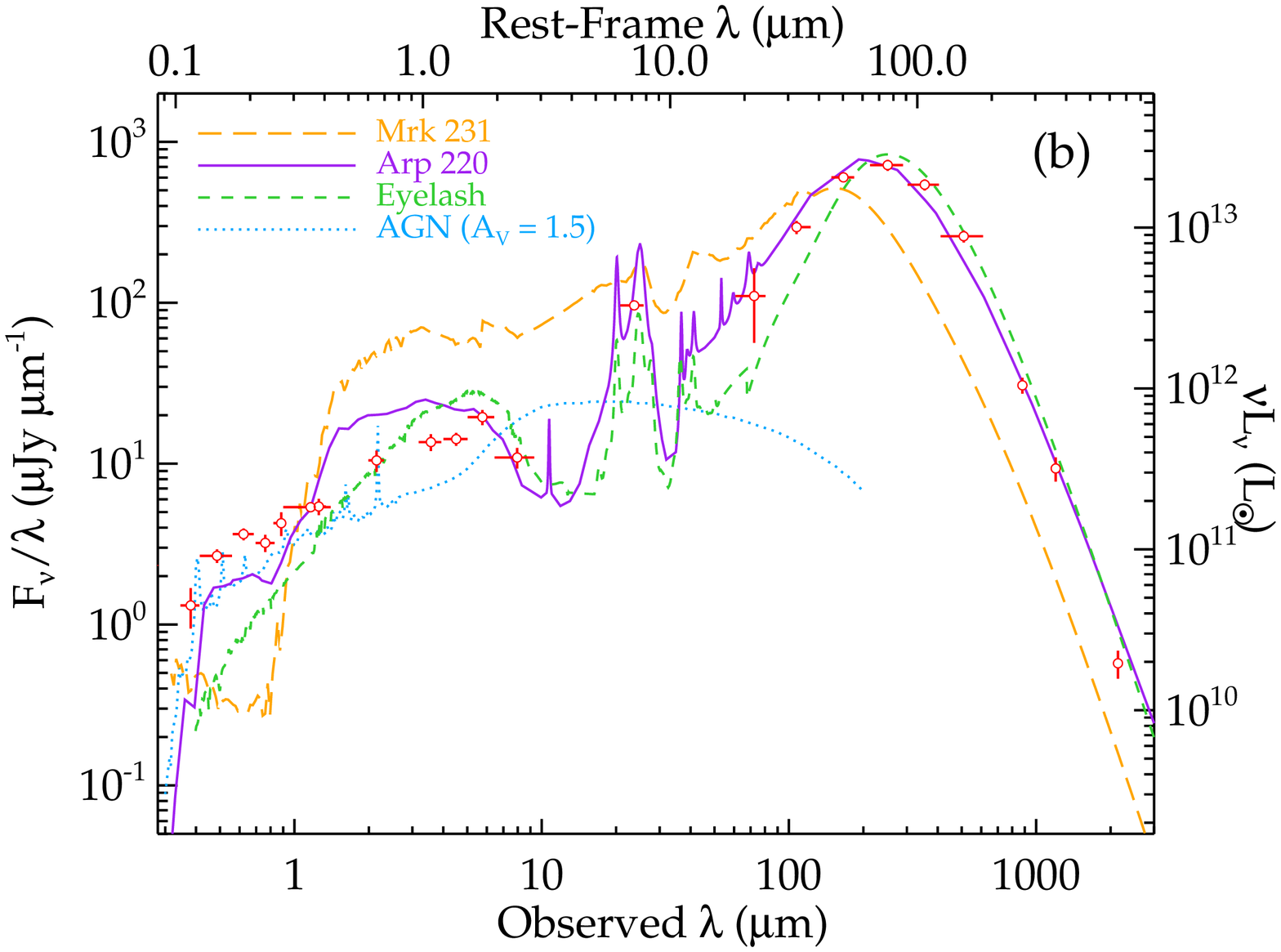}
\includegraphics[width=6.2cm]{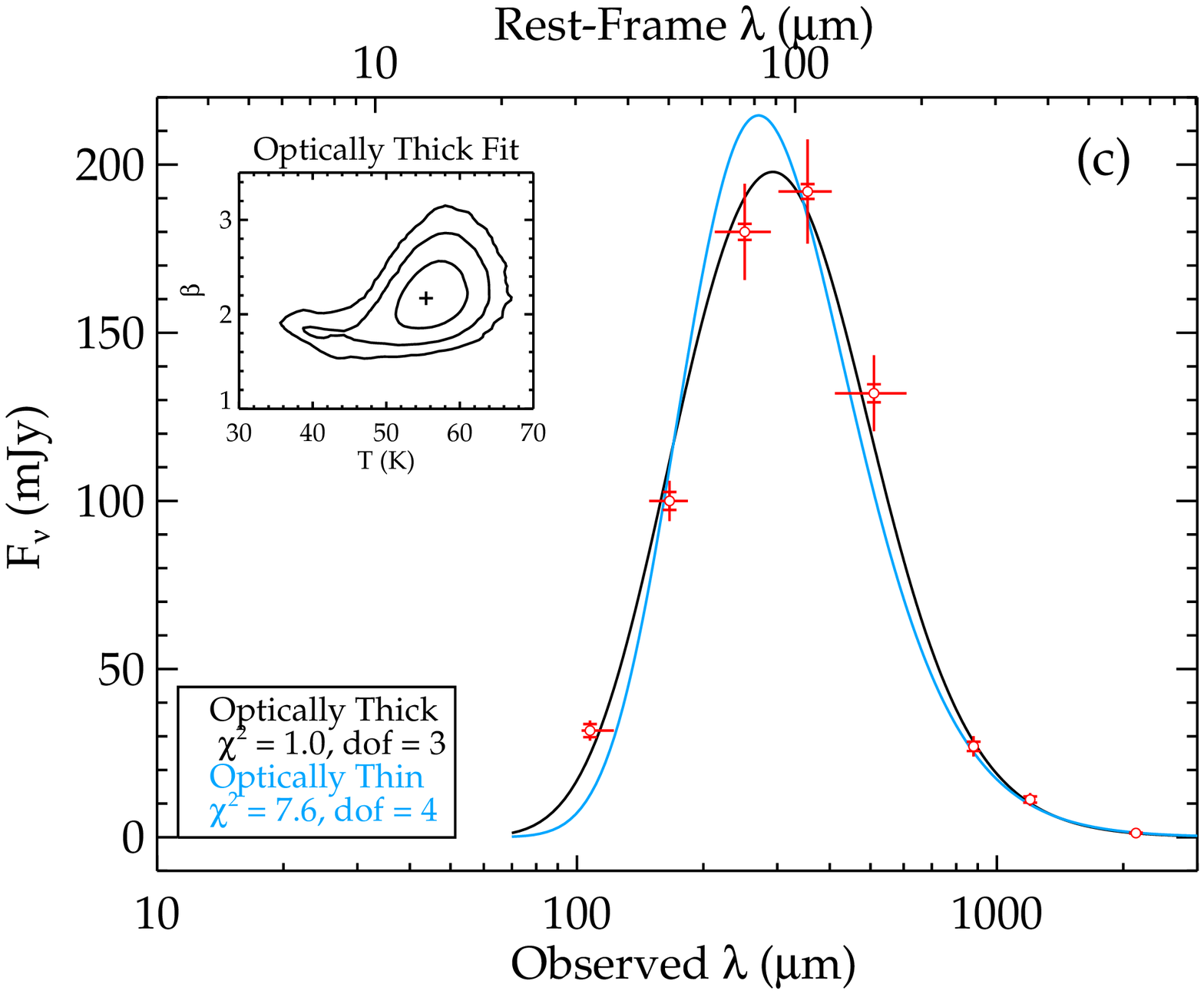}
\caption{\textbf{SED modeling of \obj}. (a) The blue and red data points are the deblended photometry of \objN\ and \objS, respectively. Their best-fit \magphys\ models (blue and red curves) are overlaid to show the quality of the fit. The dotted curves are the dust-free intrinsic (i.e., stellar only) \magphys\ models.
(b) The red data points show the total photometry of \obj\ including the filament
(Table~\ref{tab:sedobj}). Overplotted are redshifted and scaled templates from the local ULIRGs Mrk~231\cite{Chary01} (orange long dashed curve) and Arp\,220\cite{Silva98} (solid purple), the strongly lensed star-forming galaxy ``Cosmic Eyelash''\cite{Swinbank10b} at $z = 2.3$ (green dashed), and the AGN composite SED\cite{Hopkins07} (blue dotted). The AGN template has been scaled to the H$\alpha$ flux of \obj\ and reddened\cite{Calzetti00} by $A_V = 1.5$ magnitude so that it is consistent with the optical SED of \obj. (c) Blackbody fit to the far-IR (160\,\um~$< \lambda <$~2.1~mm) SED. The optically thick model (black) provides a superior fit than the optically thin model (blue); the corresponding $\chi^2$ values are labelled. The smaller error bars show only the uncorrelated statistical uncertainties, while the full error bars include both confusion noise (for PACS and SPIRE) and systematic flux calibration uncertainties. The inset shows the 1, 2, and 3$\sigma$
contours in the $T - \beta$ plane for the optically thick
model.  \label{fig:sedobj}}
\end{figure*}

We modeled the full SEDs of \objN\ and \objS\ with \magphys\ in the same
fashion as for the lensing galaxies (\S~\ref{sec:sedlens}), although
there are more data points to constrain the dust re-radiated emission.
We treat the two components of \obj\ separately because of their
distinct optical-to-near-IR SEDs. As we discussed in \S~\ref{sec:deblend}, 
\objN\ and \objS\ cannot be separated in the images at wavelengths between 24\,\um\,$< \lambda <$\,1200\,\um\ except at
880\,\um, so we had to assume that the two galaxies have the same SED shape at
these wavelengths and assign their flux densities based on their proportion in the SMA image --- 34\% in \objN, 31\%
in \objS, and 35\% in the filaments.  

The best-fit SEDs are compared with the photometry in
Figure~\ref{fig:sedobj}$a$. As expected, the level of dust extinction is high, 
with a rest-frame $V$-band optical depth of $\tau_V \sim 3-4$ for young stars
in birth clouds (i.e., only $\lesssim$5\% of the $V$-band emission leaks
out). We find that the total stellar masses are around
$(1.6\pm0.1)\times10^{11}$\,\mui\,\msun\ and $(7.4\pm0.2)\times10^{10}$\,\mui\,\msun\
for \objN\ and \objS, respectively. As we noted earlier, the compact clump at
the southern end of \objS\ dominates its emission at wavelengths shorter
than the \K-band (Figs.~\ref{fig:obs} and \ref{fig:galfit}), making it appear
much bluer than \objN. The clump may also be the reason why the model for \objS\ poorly fits the IRAC data points. Assuming that the clump is not part of \objS\ and
that the intrinsic SED of \objS\ is similar to that of \objN, we
estimate a stellar mass of $(1.7\pm0.3)\times10^{11}$\,\mui\,\msun\ using the
clump-subtracted \K-band flux ($\sim$8.1~\uJy) and the mass-to-light ratio of \objN. 
In the following we adopt this latter stellar mass estimate for \objS.

By design, the mid-to-far-IR SEDs of the two components are identical.
Integrating the best-fit SEDs between rest-frame 8 and 1000\,\um, the IR
luminosities implies an instantaneous SFR of $1100\pm100$~\mui\,\msunyr\ for each of the components, when converted using the Kennicutt\cite{Kennicutt98} calibration for a Chabrier\cite{Chabrier03} initial mass function:
\begin{equation}
 {\rm SFR}/M_{\odot}~{\rm yr}^{-1} = 10^{-10}~L_{\rm IR}/L_{\odot}
\end{equation} 

\subsection{Dust Properties} \label{sec:dust}

As the far-IR SED is dominated by thermal dust emission, 
we fit the photometric data between 160\,$\leq \lambda \leq$\,2,100\,\um\ with a
modified blackbody in both the general optically thick form ($S_\nu \propto
(1-e^{-\tau})~B_\nu(T)$, where $\tau = (\nu/\nu_0)^\beta$ and $B_\nu(T)$
is the Planck function) and the optically thin form ($S_\nu \propto
\nu^\beta~B_\nu(T)$). We treat all components of \obj\ together because they are 
spatially resolved only at 880\,\um. We fit for
the characteristic dust temperature $T_{\rm dust}$, the power law slope of dust opacity $\beta$, 
the total far-IR luminosity \lir, and the wavelength $\lambda_0 \equiv c/\nu_0$ below which the emission is 
optically thick ($\tau \geq 1$). Given that the dust opacity per unit mass follows the same power 
law as the optical depth, $\kappa_d \propto \nu^\beta$, and a normalization\cite{Dunne00,James02} of $\kappa_d(\rm 125 \mu m) = 2.64\pm0.29$~m$^2$~kg$^{-1}$, we can also estimate the dust mass $M_{\rm dust}$ assuming all dusts are at a single temperature $T_{\rm dust}$.

We use an affine-invariant Markov Chain Monte Carlo (MCMC) sampler (emcee)\cite{Foreman-Mackey12} to compute 500,000 steps from 250 samplers. Our approach fully marginalizes over the optical depth, and takes covariances between the input photometry into account. The autocorrelation length is $<$50 steps for all parameters, i.e., the chains converge very well. The resulting constraints are shown in Figure~\ref{fig:sedobj}$c$.

The general optically thick model provides a better fit to the data. Our best fits have $\chi^2 = 1$ for 3 degrees of freedom and $\chi^2 = 7.6$ for 4 degrees of freedom for the optically thick and the optically thin models, respectively. The former is therefore preferred and the best-fit parameters are tabulated in Table~\ref{tab:prop}. The total IR luminosity of \lir\,$= (3.2\pm0.2)\times10^{13}$\,\mui~\lsun\ is enormous. This implies a FIR-derived SFR of $3200\pm200$~\mui~\msunyr.

The best-fit dust temperature is $\sim$55\,K, which is much higher than the average 35~K for normal star-forming galaxies and lies at the higher end of the dust temperature distribution\cite{Hwang10} of hyper-luminous IR galaxies at $z \sim 2$. This temperature is higher than the previously derived value ($T_{\rm dust} = 43\pm1$~K)\cite{Wardlow13}, because Wardlow et al. used the optically thin model with $\beta$ fixed to 1.5, which has been ruled out because of the new PACS photometry. As a reference, our optically thin model with $\beta$ relaxed gives $T_{\rm dust} = 36\pm2$~K. 

High dust temperatures indicate efficient star formation\cite{Magnelli12b,Hayward12b}, i.e., starbursts. Using the Stephen-Boltzmann law, we have
\begin{equation}
T_{\rm dust}^{4+\beta} \propto L_{\rm IR}/M_{\rm dust} \propto {\rm SFR}/(Z_{\rm gas} M_{\rm gas}) = {\rm SFE}/Z_{\rm gas},
\end{equation}
where SFE~$\equiv$~SFR/\mg\ is the global star formation efficiency. Combined with the lack of AGN activity in \obj\ (\S~\ref{sec:AGN}), the high dust temperature clearly indicates that \obj\ is a starburst galaxy. This conclusion is supported by the star formation efficiencies per dynamical timescale estimated in \S~\ref{sec:sfe}. 

\subsection{Sizes of Star Forming Regions and Gas Reservoirs} \label{sec:size}

Our high resolution data
allow us to directly measure the extent of the dusty star forming
regions at rest frame 265\,\um\ and the gas reservoirs probed by \co.
As both components of \obj\ are resolved in the SMA and JVLA \co\
images, we use Gaussian models convolved with the beam to fit the
intensity maps. The source sizes are estimated using the best-fit
HWHMs along the major ($a$) and minor ($b$) axes ($A = \pi a b$). We
find that the total size of the dusty star forming regions ($\Sigma A_{880}
\sim 20$\,\mui\,kpc$^2$) is about five times smaller than that of the \co\ gas 
reservoirs ($\Sigma A_{\rm CO} = 95$\,\mui\,kpc$^2$),
similar to the lensed SMG HATLAS12$-$00 that has high-resolution \co\ and dust
maps\cite{Fu12b}. 

Since it is blackbody radiation, we can also estimate the cross-section of the dust emitting region using the Stefan-Boltzmann law:
\begin{equation}
A_{\rm SB} = \frac{L_{\rm IR}}{4 \sigma T_{\rm dust}^4} = 1.8\times10^{-6}~{\rm kpc}^2~L_{\rm IR} T_{\rm dust}^{-4}
\end{equation}
where $L_{\rm IR}$ is in \lsun\ and $T_{\rm dust}$ is in K. We estimate a dust
cross section of $A_{\rm SB} = 6.3\pm1.5$\,\mui\,kpc$^2$ using the optically thick model.  
This cross section based on Stefan-Boltzmann law is three times smaller than the total area measured directly from the 880~\um\ image, suggesting that the dust filling factor is only about 30\%. In other words, a lot of structures have not been resolved, consistent with the low CO brightness temperatures ($T_b = 2-3$~K) that we measured from the JVLA \co\ data cube.

\subsection{\cohh\ Conversion Factor} \label{sec:alpha}

Assuming the dust mass from the optically thick case, $M_{\rm dust} = (2.9\pm0.7)\times10^9$~\mui~\msun, and the Milky Way gas-to-dust mass ratio\cite{Draine07} of 140, we would expect a total gas mass of
$(4.1\pm1.0)\times10^{11}$\,\mui\,\msun. When compared with the \co\
luminosity from the JVLA ($L'_{\rm CO} =
(4.5\pm0.8)\times10^{11}$\,\mui\,K\,\kms\,pc$^2$), we obtain a \cohh\
conversion factor of $\alpha_{\rm CO,dust} =
0.91\pm0.27$\,\msun/(K\,\kms\,pc$^2$).  

On the other hand, through hydrodynamical simulations coupled with radiation transfer codes, the
mean velocity-integrated CO brightness temperature ($W_{\rm CO}$) from
spatially resolved \co\ data and gas metallicity ($Z/Z_{\odot}$) has
been used to provide a calibration to estimate $\alpha_{\rm
CO}$\cite{Narayanan12a}: 
\begin{equation} 
\alpha_{\rm CO,N12} = 10.7 W_{\rm CO}^{-0.32} (Z/Z_{\odot})^{-0.65} 
\end{equation}
where $W_{\rm CO} = L'_{\rm CO}/A_{\rm CO}$ is in K\,\kms, and the
resulting $\alpha_{\rm CO,N12}$ is in \msun/(K\,\kms\,pc$^2$). Using the
sizes measured from the JVLA \co\ map and assuming that $Z = Z_{\odot}$
for SMGs based on previous measurements of gas metallicity from
metallicity-diagnostic optical line ratios\cite{Swinbank04}, we estimate
that $\alpha_{\rm CO,N12} = 0.71\pm0.04$ and $0.88\pm0.04$ for \objN\ and \objS,
respectively. Note that these estimate are unaffected by lensing, 
because lensing preserves the brightness temperature.

Both of the above methods yield results consistent with the normally
assumed value of $\alpha_{\rm CO} \sim 0.8-1.0$ for mergers\cite{Downes98,Magdis11,Hodge12,Magnelli12}. 
Such a conversion factor implies tremendous amount of molecular gas and very
high gas-to-baryon fractions for both components of \obj\ (Figs.~\ref{fig:lirlco} and \ref{fig:fgas}). 

\subsection{Star Formation Efficiency} \label{sec:sfe}

\begin{figure}[!t]
\includegraphics[width=9cm]{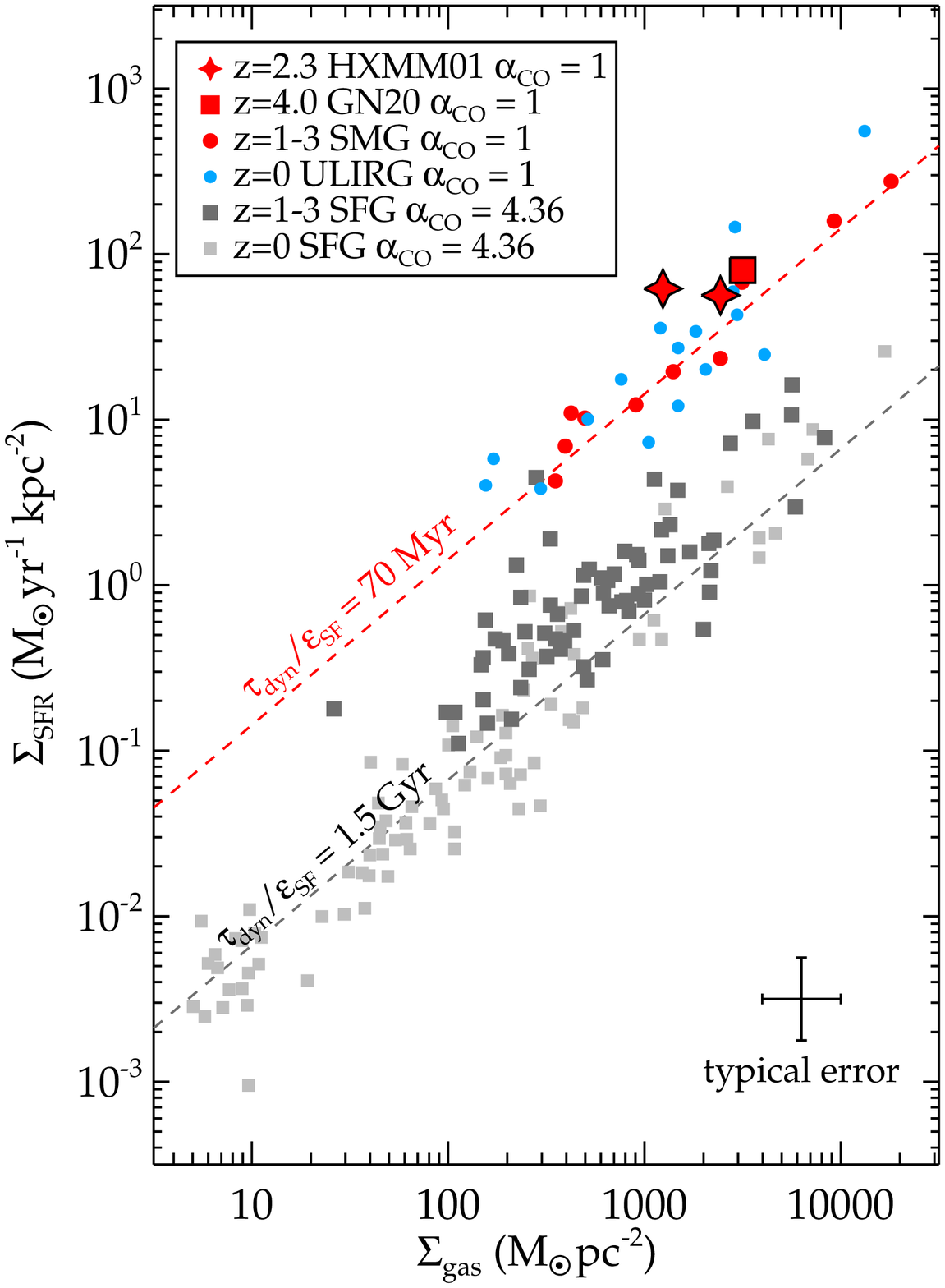}
\caption{\textbf{Star-formation-rate surface density vs. molecular gas surface
density for quiescently star-forming and starburst galaxies}. 
The two components of \obj\ and the $z = 4.05$ SMG GN20\cite{Daddi09} 
are the red stars and the red square, respectively. The 1$\sigma$ uncertainties of these measurements are smaller than the size of the symbol.
Other objects are compiled from the literature\cite{Kennicutt98b,Tacconi12}, and their typical 1$\sigma$ uncertainties are indicated by the cross at the lower right. \obj\ is consistent with the sequence defined by local\cite{Kennicutt98b} and 
high-redshift starbursts\cite{Genzel10}, which have $\sim$20 times higher SFR surface density 
at a given gas surface density than quiescently star-forming galaxies.
The dashed lines indicate constant gas consumption timescales for the star-forming disks ($\tau_{\rm disk} \equiv \tau_{\rm dyn}/\epsilon_{\rm SF} =
\Sigma_{\rm gas}/\Sigma_{\rm SFR}$) of 70~Myr (red) and 1.5~Gyr (grey). Note that $\tau_{\rm disk}$ are smaller than the consumption timescale for the entire gas reservoirs ($\tau_{\rm gas}$), because usually only part of the gas reservoir is actively forming stars.
\label{fig:sfe}} 
\end{figure}

In the form of the Kennicutt-Schmidt law that involves a dynamical
timescale\cite{Genzel10}, the star formation efficiency per dynamical timescale is
defined as
\begin{equation}
\epsilon_{\rm SF} \equiv \frac{\Sigma_{\rm SFR}}{\Sigma_{\rm gas}/\tau_{\rm dyn}}. 
\label{eq:sfe}
\end{equation}
With the measurements from the previous subsections, we can now estimate $\epsilon_{\rm SF}$ using the
SFR surface density ($\Sigma_{\rm SFR}$), the gas surface density
($\Sigma_{\rm gas}$), and the dynamical timescale ($\tau_{\rm dyn}$). 

Since the half-light radius of a two-dimensional Gaussian is equal to the
HWHM, we estimate the surface densities of the SFR and the gas mass
using the measured source sizes
\begin{equation}
\begin{array}{l}
\Sigma_{\rm SFR} = 0.5 {\rm SFR}/A_{880} ~~~\mbox{and}~~~
\Sigma_{\rm gas} = 0.5 M_{\rm H2}/A_{\rm CO},
\end{array} 
\label{eq:sigma}
\end{equation}

There are usually two dynamical timescale estimates --- the free-fall
timescale ($\tau_{\rm ff}$) and the rotational timescale ($\tau_{\rm
rot}$). We estimate the free-fall timescale as
\begin{equation}
\tau_{\rm ff} = \sqrt{R_{\rm 1/2}^3/(2 G M_{\rm vir})} = 0.74 R_{\rm 1/2}/\Delta V_{\rm FWHM},
\end{equation}
where we have used the virial mass within $R_{\rm 1/2}$ ($M_{\rm
vir}$)\cite{Spitzer87,Tacconi08}
\begin{equation}
M_{\rm vir} = 5 \sigma^2 R_{1/2}/G = 2.1\times10^5~M_\odot~\Delta V_{\rm FWHM}^2 R_{\rm 1/2}, \label{eq:mvir}
\end{equation}
where $\Delta V_{\rm FWHM}$ is in \kms\ and $R_{1/2}$ in kpc. Note that some authors\cite{Engel10,Bothwell12} used a virial
coefficient greater than 5 to account for non-virialized systems, giving a formula
$M_{\rm vir} = 2.8\times10^5~M_\odot~\Delta V_{\rm FWHM}^2 R_{\rm
1/2}$. To be consistent, we have corrected their dynamical masses to our scaling in Figure~\ref{fig:fgas}.

On the other hand, the rotational timescale is defined as
\begin{equation}
\tau_{\rm rot} = R_{\rm 1/2}/v_c = 1.36 R_{\rm 1/2}/\Delta V_{\rm FWHM},
\end{equation}
where we have used the isotropic virial formula to estimate the
characteristic circular velocity, $v_c = \sqrt{\rm{3/8ln2}} \Delta
V_{\rm FWHM} = 0.736 \Delta V_{\rm FWHM}$\cite{Tacconi12}, because the
velocity gradient is unresolved in our data. 

We opt to use the mean of the two very similar timescales as an estimate of the
dynamical timescale. For the half-light radius ($R_{1/2}$), we use the \co\ HWHM along the major
axis from the JVLA data. And for simplicity, we have considered the observed $R_{1/2}$ 
as unmagnified, although it could be magnified \emph{at most}
by a factor of $\mu$, and likely only by $\sqrt{\mu}$. Nevertheless, $\epsilon_{\rm SF}$ depends only linearly on $R_{1/2}$ because all of the other parameters involved in the equation, $\Sigma_{\rm SFR}, \Sigma_{\rm gas}$, and $\Delta V_{\rm FWHM}$, are preserved by lensing. 

With Eqn.~\ref{eq:sfe}, we estimate remarkably high
star formation efficiencies: $\epsilon_{\rm SF} \sim $10\% and 40\% for \objN\ and \objS, respectively. This is more than an order-of-magnitude higher than the mean efficiency of quiescently star-forming galaxies like our Milky Way ($\epsilon_{\rm SF} \sim $2\%). It is worth noting that the $\sim$40\% star formation efficiency of \objS\ is similar to that of the hyperluminous quasar host galaxy\cite{Walter09} SDSS~J114816.64+525150.3 at $z = 6.4$, which represents another example of maximum starbursts\cite{Murray10}. 

The SFR and gas surface densities of \obj\ are broadly consistent with the
starburst sequence defined by local ultra-luminous IR galaxies (ULIRGs) and luminous SMGs. Likewise, its SFR surface density is much higher than the quiescently star-forming galaxies at the same epoch\cite{Tacconi12} (Figure~\ref{fig:sfe}). Along with the $z=4.05$ SMG GN20\cite{Carilli10}, \obj\ seems to also show a small offset from the best-fit relation of other SMGs. This is likely because for most of the SMGs the sizes of star formation and gas reservoir were
estimated from maps of higher-$J$ CO lines\cite{Genzel10}. Lack of \co\ data, it was assumed that high-$J$ CO emission have the same physical extent as the star-forming regions and the \co\ emission, although accumulating evidence suggests that total gas reservoirs traced by \co\ emission are more extended than the star formation regions traced by submillimeter continuum and the denser gas reservoirs traced by higher-$J$ CO lines in both local ULIRGs\cite{Telesco93} and SMGs\cite{Carilli10,Ivison11,Riechers11b}. Because the higher-$J$ CO line luminosities were converted to those of \co\ using empirically determined mean brightness temperature ratios, the gas surface densities are overestimated for the SMGs that lack \co\ maps. This alluded bias is unlikely to be present in studies that have used \co\ maps for size measurements\cite{Carilli10,Ivison11,Riechers11b,Swinbank11,Fu12b}.

\subsection{Local Disk Stability} \label{sec:toomre}

As discussed in \S~\ref{sec:size}, the \co\ gaseous reservoir of \obj\ is five times larger than the dust emitting regions embedded within. The reservoir is possibly part of a massive gas inflow bringing copious amounts of cold gas to fuel the starbursts. Similar to extended CO disks in some low redshift galaxies\cite{Lucero07,Koda05,Boissier03}, it is possible that the gas reservoir of \obj\ is rotationally supported, suppressing its fragmentation to form giant molecular clouds and stars. To address the stability of the reservoir, we make use of the Toomre's criterion\cite{Toomre64}:
\begin{equation}
	Q = \frac{\kappa \sigma}{\pi G \Sigma_{\rm gas}} \label{eq:Q1}
\end{equation}
which compares the gas surface density $\Sigma_{\rm gas}$ to the critical surface density value at which the disk becomes unstable and begins to fragment. In the above, $\sigma$ is the gas velocity dispersion and $\kappa$ is the epicylic frequency, which for flat rotation curves can be written as $\kappa = \sqrt{2} v_c/R_{\rm 1/2}$, where $v_c = \sqrt{\rm{3/8ln2}} \Delta
V_{\rm FWHM} = 1.73 \sigma$ is the characteristic circular velocity\cite{Tacconi12}, and $R_{\rm 1/2}$ is the \co\ disk radius. Combined with $\Sigma_{\rm gas} = 0.5 M_{\rm H2}/(\pi R_{\rm 1/2}^2)$ and Equation~\ref{eq:mvir}, we obtain:
\begin{equation}
	Q = 0.97 \frac{M_{\rm vir}}{M_{\rm gas}} \label{eq:Q2}
\end{equation}
We estimate $Q > 1$ ($Q = 3-6$) for both \objN\ and \objS, because their virial masses are a few times greater than their gas masses. Therefore, the disks appear locally stable at least at the half-light radii. This is consistent with our finding that intense star formation occurs only in the inner parts of the disks (\S~\ref{sec:size}), since $Q$ decreases at smaller radii mainly as a result of the higher gas surface densities.

\subsection{CO Excitation}

The velocity-integrated $J$=x$\to$y to $J$=1$\to$0 brightness temperature ratio is defined as,
\begin{equation}
r_{x-y/1-1} = \frac{\int T(x-y) \, {\rm d}v}{\int T(1-0) \, {\rm d}v} =
  \frac{\int S_{\nu}(x-y) \, {\rm d}v}{\int S_{\nu}(1-0) \,
  {\rm d}v} \left( \frac{\nu_{1-0}}{\nu_{3-2}} \right)^2 \; ,
\end{equation}
where $S_\nu(x-y)$ is the flux density in the $J$=x$\to$y transition and
$\nu_{x-y}$ is the frequency of the transition. 

Combining the JVLA \co, CARMA \cothree, and PdBI \cofour\ data, we measure CO brightness temperature ratios of $r_{3-2/1-0} = 0.64\pm0.16$, and $r_{4-3/1-0} = 0.49\pm0.09$, similar to other SMGs\cite{Harris10,Ivison11,Carilli13} and $z > 2$ quasar host galaxies\cite{Riechers11a,Carilli13}. This indicates that we would have obtained a similar amount of total molecular gas mass for \obj\ 
if we had converted high-$J$ CO lines to \co\ using the mean observed brightness temperature ratios for SMGs.

\subsection{Could HXMM01 be AGN Dominated?} \label{sec:AGN}

It would be useful to know if any of our estimates are biased by the presence of AGNs.
Broad Balmer lines seen in \obj\ are common in SMGs\cite{Swinbank04,Harrison12a}. They could indicate either AGN broad
line regions or strong outflows. The clear stellar light bump sampled by the IRAC bands (Figure~\ref{fig:sedobj}$a$) indicates insignificant AGN contribution. Assuming that AGN activity is responsible for the broad H$\alpha$ lines, we can scale the composite AGN SED\cite{Hopkins07,Vanden-Berk01,Richards06} to the observed H$\alpha$ flux and compare it to the observed SED. As shown by the blue dotted curve in Figure~\ref{fig:sedobj}$b$, the AGN's hot dust bump would start to dominate the SED above $\sim$5~\um\ and produce a red [5.8]$-$[8.0] color, contradicting the observations. Dust extinction does not compromise this result: the more dust extinction we impose to the AGN SED, the redder [5.8]$-$[8.0] color we get because the AGN SED has to be scaled to the same H$\alpha$ flux. Therefore, the broad H$\alpha$ lines in \obj\ are likely produced by strong starburst-driven outflows and \obj\ is unlikely dominated by AGNs.

\subsection{X-ray Constraints on AGN luminosity and Cooling Rate of Hot Gas}\label{sec:X-ray}

Here we attempt to put some constraints on the AGN luminosity and hot
gas cooling rate using the stacked X-ray image from the XMM-LSS survey
(\S~\ref{sec:archival}), although the constraints are limited because of the shallowness of the X-ray data.

We measure the count rates with a 15\arcsec-radius aperture and the
background with an annulus between 75\arcsec\ and 90\arcsec\ (to avoid
an X-ray source 60\arcsec\ to the NE). After correcting for the aperture
loss (70\%) and the vignetting ($\sim$50\% at 10$'$
off-axis\footnote{\xmm\ Users' Handbook, \S3.2.2.2}), we measure a
3$\sigma$ upper limit of $1.1\times10^{-3}$~count~s$^{-1}$ at
0.5$-$10~keV. With
WebPIMMS\footnote{http://heasarc.gsfc.nasa.gov/Tools/w3pimms.html}, the
above count rates yield a flux of $6.6\times10^{-15}$~\ergscm\ for a
Raymond Smith plasma model of 0.2~$Z_\odot$ metallicity and a
temperature of $7.1\times10^6$~K (i.e., the virial temperature of a
$10^{13}$~\msun\ halo at $z = 2.3$), and a 0.5$-$10~keV flux of
$9.7\times10^{-15}$~\ergscm\ for a power-law model with a photon index
of 2. Therefore, the X-ray constraint on the cooling luminosity is
$L_{\rm 0.5-10~keV} < 2.8\times10^{44}$~\ergs, while the constraint on
the AGN luminosity is $L_{\rm 0.5-10~keV} < 4.2\times10^{44}$~\ergs.
These luminosities imply a cooling rate of $\dot{M} < 1,900$~\msunyr,
and an Eddington-limited black hole mass of $M_{\rm BH,Edd} \equiv
(L_{\rm bol}/1.45\times10^{38}~{\rm erg~s^{-1}})~{\rm M}_\odot <
4.8\times10^7$~\msun, assuming a bolometric correction of
6\%\cite{Elvis94}. Note that this is not an upper limit on the black
hole mass, because sub-Eddington accretion would allow higher mass black
holes.

It is not surprising that we did not detect \obj\ in the relatively
shallow X-ray observations. In the 2~Ms \chandra\ Deep Field North\cite{Alexander05}, only
two of the 20 radio-detected SMGs show X-ray fluxes greater than $f_{\rm
0.5-8~keV} > 5\times10^{-15}$~\ergscm.

\section{Space Density of Hyper-Luminous Mergers Like \obj?} \label{sec:density}

\obj\ has an intrinsic flux density of $S_{880} \simeq 17 \pm 4$~mJy. 
From the 880~\um\ single-dish source count function\cite{Karim12}, 
the surface density of sources brighter than 17~mJy at 880~\um\ is $\sim$0.1~deg$^{-2}$. 
The estimated volume density of such bright objects\cite{Karim12} is $10^{-8}$ to $10^{-7}$~Mpc$^{-3}$, much lower than volume density of SMGs with 880~\um\ flux densities of a few mJy ($\sim 10^{-6}$ to $10^{-5}$ Mpc$^{-3}$). Due to the lack of bright SMGs 
in LABOCA surveys, the volume and surface density estimates are uncertain by at least a factor of a few in either direction. Follow-up CO observations of 2 to 10~mJy SMGs find that
$16\pm13$\% (2 out of 12) are resolved\cite{Engel10} into physically related galaxy pairs with separations greater than 12~kpc. Assuming that the resolved merger fraction of fainter galaxies is applicable to more luminous sources, we estimate the surface density of bright SMG mergers like \obj\ is $\sim$0.01~deg$^{-2}$.

We can make another estimate using the source count at 500~\um. The surface density of \herschel-selected sources\cite{Wardlow13} with $S_{500} > 100$~mJy 
that are not associated with either a local ($z < 0.1$) star-forming galaxy or a bright radio-loud AGN is $0.14 \pm 0.04$~deg$^{-2}$. Follow-up high resolution imaging of ten such sources\cite{Wardlow13} reveals that \obj\ is the only well-separated SMG merger with $S_{500} > 100$~mJy in HerMES. So the expected surface density of well-resolved SMG pairs is $\sim0.014$~deg$^{-2}$, or $\sim$1 per 100~deg$^2$, consistent with our estimate from the 880~\um\ source count function. 

Since various \herschel\ surveys (HerMES, H-ATLAS, HeLMS) have mapped $\sim$1000~deg$^2$ and the South Pole Telescope\cite{Vieira10,Reichardt13} has mapped $\sim$2500~deg$^2$ in submillimeter and millimeter wavelengths, there should be $\sim$35 bright SMG mergers in these data. To identify such mergers among $\sim$480 bright lensing candidates, a good strategy is to use the CO luminosity vs. line width relation to weed out highly magnified sources (see Figure~\ref{fig:lco_fwhm}).

In Table~\ref{tab:smgmerger} we list the resolved SMG mergers reported in the literature. We note that there are many SMGs that have properties indicative of late-stage mergers. But here we only include the objects that have been resolved into multiple components. These mergers have projected separations spanning from 4 to 30~kpc and are at redshifts between $1 < z < 4$. 
References for the sources in Table~\ref{tab:smgmerger} are as follows:
SMM J02399$-$0136\citet{Valiante07,Ivison10b}; SMM J09431$+$4700\citet{Tacconi06,Engel10,Riechers11b};
SMM J105141\citet{Engel10}; SMM J123707\citet{Swinbank04,Tacconi06,Tacconi08,Engel10};
SMM J123711\citet{Chapman05,Bothwell10,Riechers11e}; SMM 163650\citet{Tacconi08}; 4C 60.07\citet{Papadopoulos00,Ivison08}.

%\clearpage

\begin{center}
\bf{\large Acknowledgement}
\end{center}

HF, AC and JLW are supported
by NASA funds for US participants in \herschel\ through an award
from JPL. 
LW and AS acknowledge support from the Science and Technology
Facilities Council [grant number ST/I000976/1]. 
AH acknowledges support from the NSF grants AST-0503946 to the University of Maryland and AST-0708653 to Rutgers University. 
MN acknowledges financial support from ASI/INAF agreement I/072/09/0.

% Keck 
Some of the data presented herein were obtained at the W.M. Keck Observatory, which is operated as a scientific partnership among the California Institute of Technology, the University of California and the National Aeronautics and Space Administration. The Observatory was made possible by the generous financial support of the W.M. Keck Foundation. The authors wish to recognize and acknowledge the very significant cultural role and reverence that the summit of Mauna Kea has always had within the indigenous Hawaiian community. We are most fortunate to have the opportunity to conduct observations from this mountain.

% CARMA
Support for CARMA construction was derived from the states of California, Illinois, and Maryland, the James S. McDonnell Foundation, the Gordon and Betty Moore Foundation, the Kenneth T. and Eileen L. Norris Foundation, the University of Chicago, the Associates of the California Institute of Technology, and the National Science Foundation. Ongoing CARMA development and operations are supported by the National Science Foundation under a cooperative agreement, and by the CARMA partner universities.

% CFHTLS
Based on observations obtained with MegaPrime/MegaCam, a joint project of CFHT and CEA/DAPNIA, at the Canada-France-Hawaii Telescope (CFHT) which is operated by the National Research Council (NRC) of Canada, the Institut National des Science de l'Univers of the Centre National de la Recherche Scientifique (CNRS) of France, and the University of Hawaii. This work is based in part on data products produced at TERAPIX and the Canadian Astronomy Data Centre as part of the Canada-France-Hawaii Telescope Legacy Survey, a collaborative project of NRC and CNRS.

% VHS
Based on observation obtained as part of the VISTA Hemisphere Survey, ESO Progam, 179.A-2010 (PI: McMahon).

% WHT
The William \herschel\ Telescope is operated on the island of La Palma by the Isaac Newton Group in the Spanish Observatorio del Roque de los Muchachos of the Instituto de Astrofsica de Canarias. The WHT observations are part of the 
International Time Programme 2010-2011 (PI Perez-Fournon).

% IRAM
Based on observations carried out with the IRAM Plateau de Bure Interferometer and 30m Telescope. IRAM is supported by INSU/CNRS (France), MPG (Germany) and IGN (Spain).

%\clearpage
%\begin{center}
%\bf{\large References}
%\end{center}

%%%%%%%%%%%%%%%%%
% Tables
%%%%%%%%%%%%%%%%%
\begin{table*}
\begin{center}
\begin{tabular}{lcccc}
\hline
\hline
Quantity & \obj & \objN & \objS & Unit \\
\hline
\multicolumn{5}{c}{Lens Modeling} \\
\hline
$\mu$(\K)    & $1.59^{+0.36}_{-0.22}$ & $1.78^{+0.55}_{-0.32}$ & $1.35^{+0.19}_{-0.12}$ & \nodata \\
$\mu$(880\um)& $1.57^{+0.34}_{-0.21}$ & $1.83^{+0.60}_{-0.35}$ & $1.34^{+0.18}_{-0.12}$ & \nodata \\
$\mu$(CO1-0) & $1.54^{+0.31}_{-0.20}$ & $1.84^{+0.63}_{-0.36}$ & $1.37^{+0.20}_{-0.13}$ & \nodata \\
Separation   & $2.22^{+0.32}_{-0.45}$/$18.7^{+2.7}_{-3.8}$ & \nd & \nd & arcsec/kpc \\
\hline
\multicolumn{5}{c}{MAGPHYS SED Modeling} \\
\hline
$M_{\rm stellar}$  & \nd & $(1.6\pm0.1)\times10^{11}$ & $(1.7\pm0.3)\times10^{11}$ & $\mu^{-1}$~\msun \\
$L_{\rm IR}$       & \nd & $(1.1\pm0.1)\times10^{13}$  & $(1.1\pm0.1)\times10^{13}$ & $\mu^{-1}$~\lsun \\
\hline
\multicolumn{5}{c}{Optically Thick Blackbody Fit} \\
\hline
$T_{\rm dust}$  & $55\pm3$ & \nd & \nd & K \\
$\beta$         & $2.17\pm0.23$ & \nd & \nd & \nodata \\
$\lambda_0$     & $180\pm30$ & \nd & \nd & \um \\
$M_{\rm dust}$  & $(2.9\pm0.7)\times10^9$ & \nd & \nd & $\mu^{-1}$~\msun \\
$L_{\rm IR}$    & $(3.2\pm0.2)\times10^{13}$ & \nd & \nd  & $\mu^{-1}$~\lsun \\
$A_{\rm SB}$    & $6.3\pm1.5$ & \nd & \nd & $\mu^{-1}$~kpc$^2$ \\
$\alpha_{\rm CO,gas/dust}$   & $0.91\pm0.27$ & \nd & \nd & \msun/(K\,\kms\,pc$^2$) \\
\hline
\multicolumn{5}{c}{SMA 880\,\um\ Image} \\
\hline
$A_{\rm 880}$            & \nd & $9.8\pm0.6$  & $8.9\pm0.7$ & $\mu^{-1}$~kpc$^2$ \\
$\Sigma_{\rm SFR}$       & \nd & $56\pm6$    & $62\pm8$     & \msunyr\,kpc$^{-2}$ \\
\hline
\multicolumn{5}{c}{JVLA CO(1$\to$0) Datacube} \\
\hline
$z_{\rm CO1-0}$             & $2.3079\pm0.0007$ & $2.3103\pm0.0007$          & $2.3074\pm0.0004$ & \nodata \\
$\Delta V_{\rm FWHM}$    & $840\pm160$       & $970\pm150$                & $660\pm100$       & \kms \\
$L'_{\rm CO1-0}$         & $(4.5\pm0.8)\times10^{11}$ & $(1.3\pm0.2)\times10^{11}$ & $(1.7\pm0.2)\times10^{11}$ & $\mu^{-1}$~K\,\kms\,pc$^2$ \\
$A_{\rm CO}$             & \nd & $27\pm2$                   & $68\pm2$ & $\mu^{-1}$~kpc$^2$ \\ 
$W_{\rm CO}$             & \nd & $4900\pm780$               & $2500\pm350$ & K\,\kms\ \\
$\alpha_{\rm CO,N12}$    & $[0.8]$ & $0.71\pm0.04$              & $0.88\pm0.04$ & \msun/(K\,\kms\,pc$^2$) \\
$M_{\rm gas}$            & $(3.6\pm0.6)\times10^{11}$ & $(9.5\pm1.0)\times10^{10}$ & $(1.5\pm0.1)\times10^{11}$ & $\mu^{-1}$~\msun \\
$\Sigma_{\rm gas}$       & \nd & $1730\pm200$               & $1090\pm100$ & \msun~pc$^{-2}$ \\
$R_{\rm 1/2}$            & \nd & $3.0\pm0.1$                & $4.6\pm0.1$ & kpc \\
$M_{\rm vir}$            & \nd & $(5.9\pm1.9)\times10^{11}$ & $(4.3\pm1.3)\times10^{11}$ & \msun \\
$\tau_{\rm rot}$         & \nd & $4.0\pm0.8$                & $9.3\pm1.5$ & Myr \\
$\tau_{\rm ff}$          & \nd & $2.2\pm0.4$                & $5.1\pm0.8$  & Myr \\
$\epsilon_{\rm SF}$      & \nd & $0.10\pm0.03$              & $0.41\pm0.10$  & \nd \\
$M_{\rm gas}/M_{\rm baryon}$   & $0.52\pm0.05$ & $0.37\pm0.04$ & $0.47\pm0.06$ \\
\hline
\multicolumn{5}{c}{GBT CO(1$\to$0) Spectrum} \\
\hline
$z_{\rm CO1-0}$          & $2.3074\pm0.0008$ & \nd & \nd & \nodata \\
$\Delta V_{\rm FWHM}$    & $1670\pm140$ & \nd & \nd  & \kms \\
$L'_{\rm CO1-0}$         & $(6.2\pm0.9)\times10^{11}$ & \nd & \nd  & $\mu^{-1}$~K\,\kms\,pc$^2$ \\
\hline
\multicolumn{5}{c}{CARMA CO(3$\to$2) Spectrum} \\
\hline
$z_{\rm CO3-2}$          & $2.3073\pm0.0010$ & \nd & \nd & \nodata \\
$\Delta V_{\rm FWHM}$    & $980\pm200$ & \nd & \nd & \kms \\
$L'_{\rm CO3-2}$         & $(2.9\pm0.5)\times10^{11}$ & \nd & \nd & $\mu^{-1}$~K\,\kms\,pc$^2$ \\
$r_{3-2/1-0}$            & $0.64\pm0.16$ & \nd & \nd & \nodata \\
\hline
\multicolumn{5}{c}{PdBI CO(4$\to$3) Spectrum} \\
\hline
$z_{\rm CO4-3}$          & $2.3081\pm0.0002$ & \nd & \nd & \nodata \\
$\Delta V_{\rm FWHM}$    & $880\pm50$ & \nd & \nd & \kms \\
$L'_{\rm CO4-3}$         & $(2.2\pm0.1)\times10^{11}$ & \nd & \nd & $\mu^{-1}$~K\,\kms\,pc$^2$ \\
$r_{4-3/1-0}$            & $0.49\pm0.09$ & \nd & \nd & \nodata \\
\hline
\multicolumn{5}{c}{Keck H$\alpha$ Spectra} \\
\hline
$z_{\rm H\alpha}$        & \nd & $2.3144\pm0.0019$ & $2.3107\pm0.0015$ & \nodata \\ 
$\Delta V_{\rm FWHM}$    & \nd & $2,370\pm340$     & $1,750\pm320$ & \kms \\
$L_{\rm H\alpha}$        & \nd & $(9.4\pm1.6)\times10^{42}$ & $(1.4\pm0.2)\times10^{43}$ & $\mu^{-1}$~erg\,s$^{-1}$ \\
\hline
\end{tabular}
\end{center}
\caption{Observed and Derived Properties of \obj. Refer to text for definitions of the parameters.}
\label{tab:prop}
\end{table*}

\begin{table*}[!b]
\begin{center}
\begin{tabular}{lcccc}
\hline
\hline
Instrument & Band & $\lambda$  & G1 & G2 \\
           &      & (\um)      & (\uJy)  & (\uJy) \\
\hline
GALEX    & NUV      &   $   0.23$ & $<0.5$ & $<0.5$ \\
CFHT     & u*       &   $   0.38$ & $     0.7\pm    0.2$ & $     1.0\pm    0.2$ \\
CFHT     & g'       &   $   0.49$ & $     2.9\pm    0.2$ & $     4.1\pm    0.2$ \\
CFHT     & r'       &   $   0.62$ & $     6.3\pm    0.3$ & $    14.7\pm    0.5$ \\
CFHT     & i'       &   $   0.76$ & $    14.8\pm    0.5$ & $    31.1\pm    1.0$ \\
CFHT     & z'       &   $   0.88$ & $    19.8\pm    0.9$ & $    46.3\pm    1.6$ \\
HST      & F110W    &   $   1.16$ & $    29.0\pm    0.8$ & $    72.9\pm    1.7$ \\
WHT      & J        &   $   1.25$ & $    32.7\pm    1.2$ & $    78.2\pm    1.8$ \\
VISTA    & H        &   $   1.65$ & $    60.2\pm    5.3$ & $   138.5\pm    6.5$ \\
WHT      & K        &   $   2.15$ & $    87.7\pm    7.7$ & $   184.6\pm    5.1$ \\
IRAC     & 3.6\,\um &   $   3.56$ & $   111.1\pm    7.9$ & $   180.6\pm    7.9$ \\
IRAC     & 4.5\,\um &   $   4.51$ & $    77.6\pm    6.2$ & $   146.5\pm    7.3$ \\
IRAC     & 5.8\,\um &   $   5.76$ & $    70.4\pm   13.9$ & $    99.3\pm   13.7$ \\
IRAC     & 8.0\,\um &   $   7.96$ & $    51.4\pm   14.6$ & $   106.8\pm   14.7$ \\
\hline
\end{tabular}
\end{center}
\caption{Photometry of the foreground lensing galaxies.}
\label{tab:sedg1g2}
\end{table*}

\begin{table*}
\begin{center}
\begin{tabular}{lccccc}
\hline
\hline
Instrument & Band & $\lambda$ & \obj & \objN & \objS \\
           &      & (\um)     & (\uJy)    & (\uJy)     & (\uJy)     \\
\hline
GALEX    & NUV      &  $    0.23$ & $<0.5$ & $<0.5$ & $<0.5$                                           \\
CFHT     & u*       &  $    0.38$ & $     0.5\pm    0.1$ & $<0.2    $           & $     0.5\pm    0.1$ \\
CFHT     & g'       &  $    0.49$ & $     1.3\pm    0.1$ & $<0.2    $           & $     1.2\pm    0.1$ \\
CFHT     & r'       &  $    0.62$ & $     2.3\pm    0.2$ & $     0.2\pm    0.1$ & $     2.1\pm    0.2$ \\
CFHT     & i'       &  $    0.76$ & $     2.5\pm    0.3$ & $<0.5    $           & $     2.3\pm    0.3$ \\
CFHT     & z'       &  $    0.88$ & $     3.8\pm    0.6$ & $     0.6\pm    0.3$ & $     3.2\pm    0.6$ \\
HST      & F110W    &  $    1.16$ & $     6.2\pm    0.4$ & $     1.6\pm    0.2$ & $     4.7\pm    0.3$ \\
WHT      & J        &  $    1.25$ & $     6.8\pm    0.8$ & $     1.7\pm    0.5$ & $     5.1\pm    0.6$ \\
VISTA    & H        &  $    1.65$ &         \nodata      & \nodata              &         \nodata      \\
WHT      & K        &  $    2.15$ & $    22.5\pm    3.4$ & $     7.6\pm    2.0$ & $    14.9\pm    2.7$ \\
IRAC     & 3.6\,\um &  $    3.56$ & $    48.5\pm    5.8$ & $    22.2\pm    5.0$ & $    26.4\pm    2.8$ \\
IRAC     & 4.5\,\um &  $    4.51$ & $    64.1\pm    5.9$ & $    29.4\pm    5.0$ & $    34.7\pm    2.9$ \\
IRAC     & 5.8\,\um &  $    5.76$ & $   111.9\pm   12.0$ & $    51.5\pm    6.8$ & $    60.4\pm    9.6$ \\
IRAC     & 8.0\,\um &  $    7.96$ & $    86.9\pm   12.7$ & $    41.1\pm    6.7$ & $    45.8\pm   10.6$ \\
MIPS     & 24\,\um  &  $   23.67$ & $  2280  \pm   100$ & \nodata              & \nodata              \\
PACS     &  70\,\um &  $   72.35$ & $  7970  \pm 3890  $ & \nd & \nd \\
PACS     & 100\,\um &  $  107.42$ & $ 31700  \pm 3000  $ & \nd & \nd \\
PACS     & 160\,\um &  $  166.15$ & $102100  \pm 6000  $ & \nd & \nd \\
SPIRE    & 250\,\um &  $  250.94$ & $180300  \pm14300  $ & \nodata              & \nodata              \\
SPIRE    & 350\,\um &  $  354.27$ & $192100  \pm15500  $ & \nodata              & \nodata              \\
SPIRE    & 500\,\um &  $  509.47$ & $131600  \pm11300  $ & \nodata              & \nodata              \\
SMA      & 880\,\um &  $  876.50$ & $ 27000  \pm 3000  $ & $  9300  \pm 1200  $ & $  8300  \pm 1100  $ \\
MAMBO    & 1.2\,mm  &  $ 1200.00$ & $ 11200  \pm 1900  $ & \nodata              & \nodata              \\
PdBI     & 2.1\,mm  &  $ 2141.31$ & $  1220  \pm 240  $  & \nodata              & \nodata              \\ 
VLA      & 1.4~GHz  & $2.1\times10^5$ & $<420  $ & $<420  $ & $<420  $                                 \\
\hline
\end{tabular}
\end{center}
\caption{Photometry of \obj.}
\label{tab:sedobj}
\end{table*}

\end{document}